\documentclass[12pt]{iopart}
\usepackage{iopams}  
\usepackage[table,xcdraw]{xcolor}
\usepackage[utf8]{inputenc}
\usepackage{color}
\usepackage{graphicx}
\usepackage[flushleft]{threeparttable}
\usepackage{multicol}
\usepackage{upgreek}
\usepackage[numbers,sort&compress,square]{natbib}

\newcommand{\Ms}{M_\odot}
\newcommand{\eqref}[1]{(\ref{#1})}

\begin{document}

\title[The missing link in GW astronomy]{The missing link in gravitational-wave astronomy: Discoveries waiting in the decihertz range}

\author[1]{Manuel Arca Sedda$^1$, Christopher P L Berry$^{2,3}$, Karan Jani$^4$, Pau Amaro-Seoane$^{5,6,7,8}$, Pierre Auclair$^9$, Jonathon Baird$^{10}$, Tessa Baker$^{11}$, Emanuele Berti$^{12}$, Katelyn Breivik$^{13}$, Adam Burrows$^{14}$, Chiara Caprini$^9$, Xian Chen$^{15,6}$, Daniela Doneva$^{16}$, Jose M Ezquiaga$^{17}$, K E Saavik Ford$^{18}$, Michael L Katz$^2$, Shimon Kolkowitz$^{19}$, Barry McKernan$^{18}$, Guido Mueller$^{20}$, Germano Nardini$^{21}$, Igor Pikovski$^{22}$, Surjeet Rajendran$^{12}$, Alberto Sesana$^{23}$, Lijing Shao$^{6}$, Nicola Tamanini$^{24}$, David Vartanyan$^{25}$, Niels Warburton$^{26}$, Helvi Witek$^{27}$, Kaze Wong$^{12}$, Michael Zevin$^2$}
\address{$^1$ Astronomisches Rechen-Institut, Zentr\"um f\"ur Astronomie, Universit\"at Heidelberg, M\"onchofstr. 12-14, Heidelberg, Germany}
\address{$^2$ Center for Interdisciplinary Exploration and Research in Astrophysics (CIERA),  Department of Physics and Astronomy, Northwestern University, 1800 Sherman Avenue, Evanston, IL 60201, USA}
\address{$^3$ SUPA, School of Physics and Astronomy, University of Glasgow, Glasgow G12 8QQ, UK}
\address{$^4$ Department of Physics and Astronomy, Vanderbilt University, Nashville, TN 37212, USA }
\address{$^5$ Universitat Polit{\`e}cnica de Val{\`e}ncia, IGIC, Spain}
\address{$^6$ Kavli Institute for Astronomy and Astrophysics, Peking University, Beijing 100871, China}
\address{$^7$ Institute of Applied Mathematics, Academy of Mathematics and Systems Science, CAS, Beijing 100190, China}
\address{$^8$ Zentrum f\"{u}r Astronomie und Astrophysik, TU Berlin, Hardenbergstraße 36, 10623 Berlin, Germany}
\address{$^9$ Laboratoire Astroparticule et Cosmologie, CNRS UMR 7164, Universit\'{e} Paris-Diderot, 10 rue Alice Domon et L\'{e}onie Duquet, 75013 Paris, France}
\address{$^{10}$ High Energy Physics Group, Physics Department, Imperial College London, Blackett
Laboratory, Prince Consort Road, London, SW7 2BW, UK}
\address{$^{11}$ School of Physics and Astronomy, Queen Mary University of London, Mile End Road, London, E1 4NS, UK}
\address{$^{12}$ Department of Physics and Astronomy, Johns Hopkins University, 3400 N.\ Charles Street, Baltimore, MD 21218, USA}
\address{$^{13}$ Canadian Institute for Theoretical Astrophysics, University of Toronto, 60 St.\ George Street, Toronto, Ontario, M5S 1A7, Canada}
\address{$^{14}$ Department of Astrophysical Sciences, University of Princeton, 4 Ivy Lane, Princeton, NJ 08544-1001, US}
\address{$^{15}$ Astronomy Department, School of Physics, Peking University, Beijing 100871, China}
\address{$^{16}$ Theoretical Astrophysics, Eberhard Karls University of T\"ubingen, T\"ubingen 72076, Germany}
\address{$^{17}$ Kavli Institute for Cosmological Physics and Enrico Fermi Institute, The University of Chicago, Chicago, IL 60637, USA}
\address{$^{18}$ City University of New York-BMCC, 199 Chambers St, New York, NY 10007, USA \& Dept. of Astrophysics, American Museum of Natural History, New York, NY 10028, USA}
\address{$^{19}$ Department of Physics, University of Wisconsin -- Madison, Madison, WI 53706, USA}
\address{$^{20}$ Department of Physics, University of Florida, PO Box 118440, Gainesville, Florida 32611, USA}
\address{$^{21}$ Faculty of Science and Technology, University of Stavanger, 4036 Stavanger, Norway}
\address{$^{22}$ Department of Physics, Stevens Institute of Technology, Hoboken, NJ 07030, USA \& The Oskar Klein Centre, Department of Physics, Stockholm University, Stockholm, Sweden}
\address{$^{23}$ Universit\`a di Milano Bicocca, Dipartimento di Fisica G.\ Occhialini, Piazza della Scienza 3, I-20126, Milano, Italy}
\address{$^{24}$ Max-Planck-Institut für Gravitationsphysik (Albert-Einstein-Institut), Am Mühlenberg 1, 14476 Potsdam-Golm, Germany}
\address{$^{25}$ Department of Astronomy (University of California), Berkeley, CA 94720-3411, USA}
\address{$^{26}$ School of Mathematics and Statistics, University College Dublin, Belfield, Dublin 4, Ireland}
\address{$^{27}$ Department of Physics, King's College London, Strand, London, WC2R 2LS, United Kingdom}

\eads{\mailto{m.arcasedda@uni-heidelberg.de}, \mailto{christopher.berry@northwestern.edu}, \mailto{karan.jani@vanderbilt.edu}}
\vspace{10pt}
\begin{indented}
\item[]\today
\end{indented}

\begin{abstract}
    The gravitational-wave astronomical revolution began in 2015 with LIGO's observation of the coalescence of two stellar-mass black holes. 
    Over the coming decades, ground-based detectors like LIGO, Virgo and KAGRA will extend their reach, discovering thousands of stellar-mass binaries. 
    In the 2030s, the space-based \textit{LISA} will enable gravitational-wave observations of the massive black holes in galactic centres. 
    Between ground-based observatories and \textit{LISA} lies the unexplored decihertz gravitational-wave frequency band. 
    Here, we show the potential of a \emph{Decihertz Observatory} which could cover this band, and complement discoveries made by other gravitational-wave observatories. 
    The decihertz range is uniquely suited to observation of intermediate-mass ($\sim10^2$--$10^4 \Ms$) black holes, which may form the missing link between stellar-mass and massive black holes, offering an opportunity to measure their properties.  
    Decihertz observations will be able to detect stellar-mass binaries days to years before they merge and are observed by ground-based detectors, providing early warning of nearby binary neutron star mergers, and enabling measurements of the eccentricity of binary black holes, providing revealing insights into their formation. 
    Observing decihertz gravitational-waves also opens the possibility of testing fundamental physics in a new laboratory, permitting unique tests of general relativity and the Standard Model of particle physics.
    Overall, a Decihertz Observatory would answer outstanding questions about how black holes form and evolve across cosmic time, open new avenues for multimessenger astronomy, and advance our understanding of gravitation, particle physics and cosmology.
\end{abstract}

\vspace{2pc}

\submitto{\CQG}

\section{The astronomical revolution}

During the 20th century there was an explosion of astronomical discoveries as new instruments enabled us to observe more of the electromagnetic spectrum~\citep{Longair:2006}. 
Diversifying out from visible light provided a richer understanding of the cosmos and provided many unexpected discoveries---from radio pulsars~\citep{Hewish:1968bj,Beskin:2015ria} to gamma-ray bursts~\citep{Klebesadel:1973iq,Meszaros:2006rc,Berger:2013jza}. 
A similar revolution awaits gravitational-wave (GW) astronomy~\citep{Sathyaprakash:2009xs,Abbott:2016blz}, and we discuss the {scientific potential of exploring the ${\sim0.01}$--${1~\mathrm{{Hz}}}$ GW spectrum}.

On 14 September 2015 the twin Laser Interferometer Gravitational-Wave Observatory (LIGO) detectors made the first observation of a GW signal~\citep{Abbott:2016blz}. 
GW150914 originated from a coalescence of two black holes (BHs) each about $30 \Ms$~\citep{TheLIGOScientific:2016wfe,LIGOScientific:2018mvr}. 
This discovery enabled revolutionary advances in the understanding of the astrophysics of binary BHs (BBHs)~\citep{TheLIGOScientific:2016htt,LIGOScientific:2018jsj} and the nature of gravity~\citep{TheLIGOScientific:2016src,LIGOScientific:2019fpa}. 
The signal was observed sweeping through a frequency range of $\sim20$--$250~\mathrm{Hz}$. 
The lower frequency limit is set by the sensitivity of the detectors, because seismic noise prevents observations at low frequencies. 
The upper limit is set by the merger frequency of the BHs, which is inversely proportional to the binary's total mass. 
Ground-based detectors, like LIGO~\citep{TheLIGOScientific:2014jea}, Virgo~\citep{TheVirgo:2014hva} and KAGRA~\citep{Akutsu:2018axf}, can observe across a range of frequencies $\sim10$--$10^3~\mathrm{Hz}$. 
This is well tailored to the detection of merging stellar-mass BH and neutron star (NS) binaries~\citep{LIGOScientific:2018mvr,Abbott:2020uma,LIGOScientific:2020stg,Abbott:2020khf}, but it is only a small part of the GW spectrum. 
Next-generation ground-based detectors, like Cosmic Explorer~\citep{Evans:2016mbw} or the Einstein Telescope~\citep{Sathyaprakash:2012jk} may extend the observable range of GW frequencies down to a few hertz, but pushing lower requires switching to space-based observatories.

Observing at lower GW frequencies enables study of the mergers of more massive binaries, and measurements of stellar-mass binaries earlier in their inspirals. 
The space-based 
\textit{Laser Interferometer Space Antenna} (\textit{LISA}), due for launch in 2034, will observe across frequencies $\sim10^{-4}$--$10^{-1}~\mathrm{Hz}$, being most sensitive around $3\times10^{-3}~\mathrm{Hz}$~\citep{Audley:2017drz}. 
This makes it perfectly suited to observe the merger of binaries with $\sim10^6 \Ms$ massive BHs~\citep{Klein:2015hvg,Babak:2017tow,Berry:2019wgg}. 
These massive BHs are found in the centres of galaxies~\citep{Kormendy:1995er,Ferrarese:2004qr}, including our own Milky Way~\citep{Abuter:2018drb}. 
\textit{LISA} would also have been able to observe a binary like GW150914's source years--days prior to merger~\citep{Sesana:2016ljz}. 
Making \emph{multiband} observations of stellar-mass binaries opens up new avenues of investigation, including unravelling how the systems formed~\citep{Breivik:2016ddj,Nishizawa:2016jji} and enabling precision tests of gravity~\citep{Vitale:2016rfr,Liu:2020nwz}.

At even lower frequencies, pulsar timing arrays are sensitive to GWs of $\sim10^{-9}$--$10^{-7}~\mathrm{Hz}$~\citep{IPTA):2013lea}. 
This makes them well suited to observe $\sim10^9 \Ms$ supermassive BHs~\citep{Mingarelli:2017fbe}. 
How (super)massive BHs form and evolve is currently an active area of research with many outstanding questions. 
Combining observations from \textit{LISA} and pulsar timing will provide a unique insight into the growth of supermassive BHs~\citep{Pitkin:2008iu,Colpi:2019yzd}.

Here, we explore the case for extending the accessible GW spectrum with an observatory that can observe in the $\sim 0.01$--$1~\mathrm{Hz}$ \emph{decihertz} range. 
Such observations would (i) unravel the channels driving the formation of stellar-mass binaries, enhancing ground-based observations with deeper multiband observations compared to those achievable with \textit{LISA}; 
(ii) complete our picture of the population of BHs by providing unrivaled measurements of intermediate-mass BHs (IMBHs), which may be the missing link in the formation and evolution of (super)massive BHs, and (iii) enable tests of fundamental physics in a new regime. 
A decihertz GW observatory has the unrivaled potential to answer questions about the complicated physical processes that regulate binary star formation and evolution, to examine the formation of astrophysical BHs at all scales back to the early Universe, and to look for the existence of extensions to general relativity (GR) or the Standard Model.

\section{Opening the decihertz window}

\begin{figure}[ht!]
\centering
 \includegraphics[scale=0.6, trim = {-30 0 0 0}]{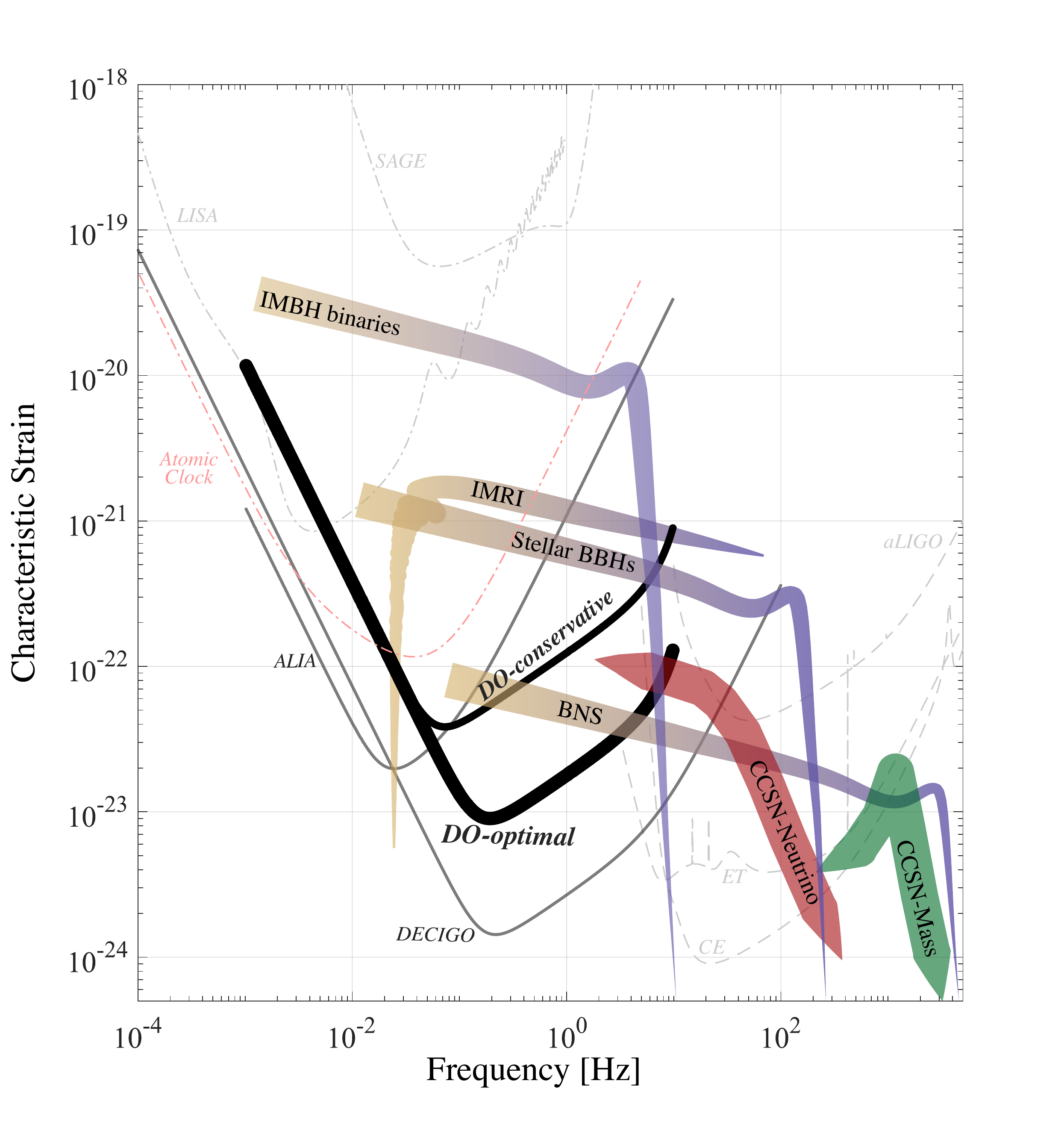} 
\caption{Evolution of binaries from millihertz to decihertz to hectohertz, together with sensitivity curves for space-based and ground-based gravitational-wave observatories. 
Concept designs for Decihertz Observatories (DOs) fill the gap between \textit{LISA} and ground-based detectors like Advanced LIGO (aLIGO), Cosmic Explorer and the Einstein Telescope.  
Details about different design curves and corresponding parameters are discussed in Section~\ref{Sec:Detector}.
}
\label{Fig1}
\end{figure}

In the following sections, we will highlight some of the scientific opportunities of $0.01$--$1~\mathrm{Hz}$ GW measurements. 
Employing decihertz observations would bridge space-based low-frequency detectors, like \textit{LISA}, and ground-based experiments, like LIGO and Virgo or their next-generation successors Cosmic Explorer~\citep{Evans:2016mbw} or the Einstein Telescope~\citep{Sathyaprakash:2012jk}. 
Spanning this GW frequency spectrum will open new scientific highways, making possible the detection of new sources and phenomena, as well as enhancing our understanding of multiband sources. 
Among the population of astrophysical systems available to study are:
\begin{enumerate}
    \item Stellar-mass binaries of compact stellar remnants---white dwarfs (WDs), NSs and stellar-mass BHs (Section~\ref{Sec:CBC}). 
    BH and NS mergers are observable with ground-based detectors, hence it is possible to have combined multiband observations of these populations, with decihertz observations providing valuable forewarning of mergers.
    Decihertz observations of mergers can provide critical forewarning of multimessenger emission associated with merger events.
    Following their detection by LIGO and Virgo, BHs and NSs are a \emph{guaranteed} class of GW source~\citep{Abbott:2016blz,LIGOScientific:2018mvr,Abbott:2020uma,LIGOScientific:2020stg,Abbott:2020khf}.
    \item The elusive IMBHs of $\sim10^2$--$10^4 \Ms$. 
    The decihertz GW range is perfectly suited to study BHs in this mass range.  
    IMBHs could be discovered in binaries with compact stellar remnants as intermediate mass-ratio inspirals (IMRIs), or in a coalescing binary with another IMBH (Section~\ref{Sec:IMBH}).
    \item Cosmological sources as part of a stochastic GW background (SGWB). 
    Both this, and the other astrophysical sources serve as probes of new physics, enabling tests of modifications to GR and the Standard Model of particle physics (Section~\ref{Sec:fun}). 
    Observing a modification to either of these cornerstones of modern physics would revolutionize our understanding of the Universe, whereas recording no deviations would place new, stringent bounds on viable alternative theories.
\end{enumerate}
Decihertz observations provide a unique insight into the physics of each of these sources, and observations would answer questions on diverse topics ranging from the dynamics of globular clusters to the nature of dark matter.

The scientific return for each of the source classes is dependent upon the observatory design. 
There are multiple potential technologies and mission designs for observing the $0.01$--$1~\mathrm{Hz}$ GW spectrum. 
We will refer to this class of detectors as \emph{Decihertz Observatories} (DOs). 
To illustrate the potential of DOs, we pick illustrative configurations to highlight what would be possible with an observatory of comparable sensitivity. 
Our ensemble of illustrative detectors consists of: two illustrative \textit{LISA}-like designs, the more ambitious DO-Optimal and the less challenging DO-Conservative, and two DO concepts currently in the literature, the \textit{Advanced Laser Interferometer Antenna} (\textit{ALIA})~\citep{Bender:2013nsa,Baker:2019pnp} and the \textit{DECi-hertz Interferometer Gravitational-wave Observatory} (\textit{DECIGO})~\citep{Sato:2017dkf,Kawamura:2020pcg}.  
Together, these designs span a range of sensitivities across the decihertz range. 
Comparing the scientific capabilities of our illustrative designs gives an indication of the potential cost/benefit trade-offs between potential mission designs.
Details of these designs and the current technological readiness are reviewed in Section~\ref{Sec:Detector}.
The projected sensitivities of DO concepts are illustrated in Figure~\ref{Fig1}; to highlight the potential of these detectors, we overlay signals associated with different sources: NS binaries (component masses $1.4\Ms+1.4\Ms$), stellar-mass BBHs ($30\Ms+30\Ms$), IMRIs ($10^3\Ms+10\Ms$) with eccentricity $0.9$, and IMBH binaries ($10^3\Ms+10^3\Ms$).

\begin{figure}[ht!]
\centering
 \includegraphics[width=\textwidth,trim = {0 0 0 0}]{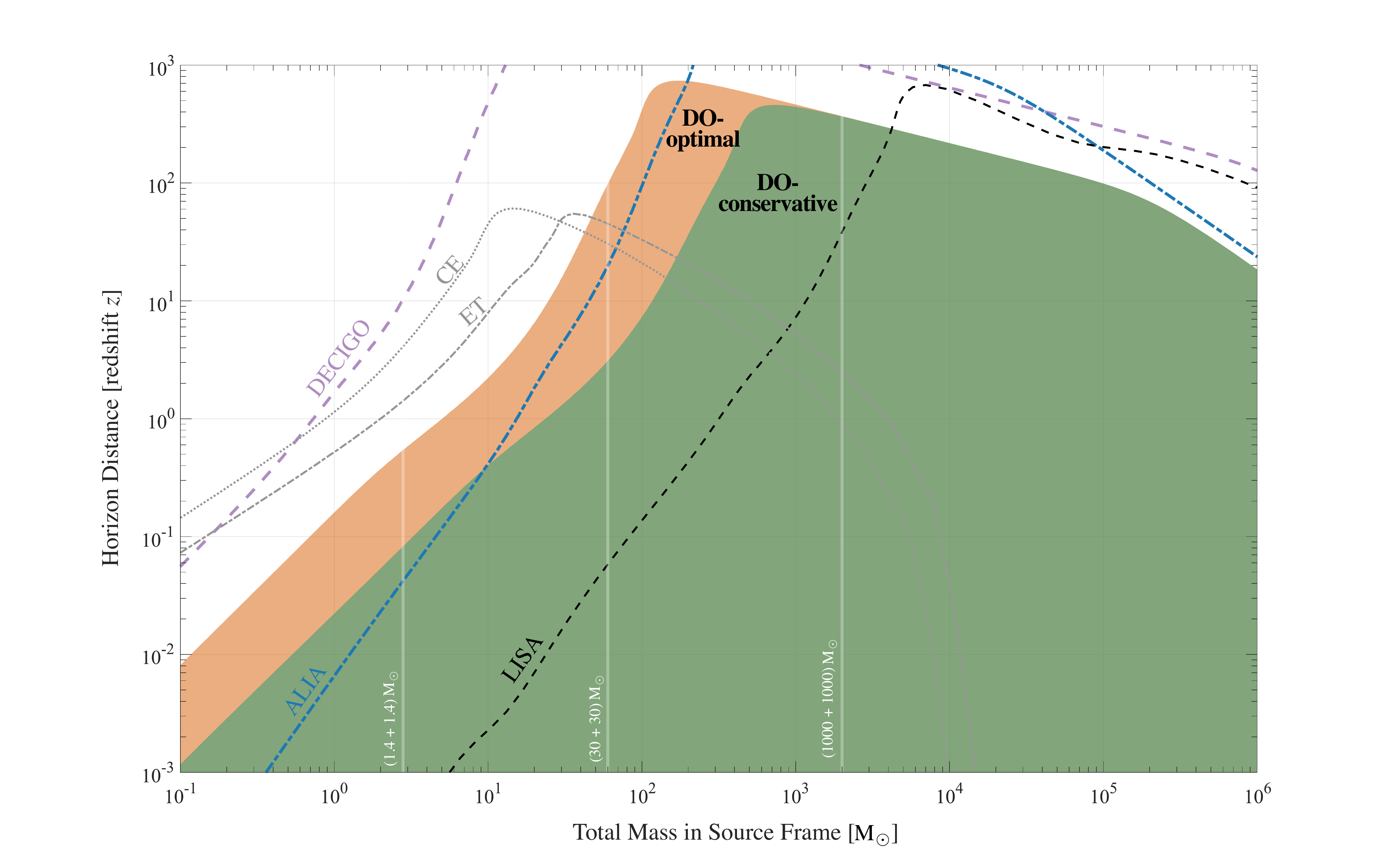} 
\caption{Cosmological reach to black hole and neutron star binaries in the proposed Decihertz Observatories (DOs). 
Binaries containing objects less compact than black holes or neutron stars, such as white dwarfs (see Sec.~\ref{Sec:WD}) or main-sequence stars, merge at wider separations corresponding to lower gravitational-wave frequencies, and so do not have the same detection ranges. 
Binaries are assumed to be circular, and components are assumed to be equal mass and have zero spin.}.
\label{Fig2}
\end{figure}

Figure~\ref{Fig1} shows the potential of decihertz observatories in bridging low- and high-frequency detectors. 
The capabilities of a GW observatory can be quantified using the horizon redshift, the maximum distance at which a given type of source can be detected assuming a threshold signal-to-noise ratio (SNR). 
Figure~\ref{Fig2} shows the horizon redshift for BH and NS binaries with total masses in the range $0.1$--$10^6 \Ms$, assuming an SNR threshold of $12$. 
The DO-Optimal concept would enable us to study BH pairs with masses in the range $\sim 30$--$10^3 \Ms$ up to the dawn of the first stars, providing deeper observations the Einstein Telescope or \textit{LISA}. 
Such observations would enable exploration of connections between the first stars, stellar BHs, IMBHs and the growth of SMBHs. 
Table~\ref{tab:1} summarizes the horizon redshift calculated for several types of GW sources (assuming circular binaries with zero spin components) and different DO designs.

%%%%%%%%%%%%%%%%%%%%%%%%%%%%%%%%%%%%%%%%%%%%%%%%%%%%%%%%%%%%%%%%%%%%%%%%%%%

\section{Stellar-mass binaries \& multiband gravitational-wave astronomy}
\label{Sec:CBC}

\begin{table}[]
\begin{center}
\caption{Horizon redshift for different sources and gravitational-wave observatories. 
Binaries are assumed to be circular, and components are assumed to have zero spin.}
\label{tab:Horizons}
\begin{tabular}{lcccccc}
\br
 \multicolumn{1}{c}{Source type} & \multicolumn{1}{c}{Masses $(m_1+m_2)/\Ms$} & \multicolumn{1}{c}{DO-Conservative} & \multicolumn{1}{c}{DO-Optimal} & \multicolumn{1}{c}{\textit{ALIA}} & \multicolumn{1}{c}{\textit{DECIGO}} \\
\mr
BNS             & $1.4 + 1.4$   &  $0.1$    & $ 0.5$  & $ 0.05$   & $ 10$       \\
NSBH            & $14 + 1.4$    &  $0.3$    & $ 2$    & $ 0.5$    & $ 170$      \\
NSBH            & $70 + 1.4$    &  $0.6$    & $ 4$    & $ 2$      & $ 250$      \\
BBH             & $30 + 30$     &  $3$      & $ 100$  & $ 20$     & $>10^3$     \\
IMRI (IMBH--BH) & $100 + 10$    &  $3$      & $ 60$   & $ 20$     & $>10^3$     \\
IMRI (IMBH--BH) & $2000 + 40$   &  $30$     & $ 50$   & $ 180$    & $ 160$      \\
IMBH--IMBH      & $1000 + 100$  &  $90$     & $ 160$  & $ 600$    & $ 470$      \\
IMBH--IMBH      & $1000 + 1000$ &  $370$    & $ 380$  & $>10^3$   & $>10^3$     \\
\br
\end{tabular}
\label{tab:1}
    \end{center}
\end{table}

\subsection{Revealing binary evolution channels}

LIGO and Virgo have opened the window to the GW Universe with their observations of a stellar-mass BBH~\citep{Abbott:2016blz}. 
Following the completion of their second observing run, $10$ BBHs and $1$ binary NS (BNS) have been observed~\citep{LIGOScientific:2018mvr};  
data from their third observing run are still being analysed, but already $1$ BNS~\citep{Abbott:2020uma}, $1$ BBH~\citep{LIGOScientific:2020stg}, and one system which could be either a NS--BH (NSBH) binary or a BBH~\citep{Abbott:2020khf} have been announced. 
With measurements of these systems' masses and spins we can start to reconstruct the characteristics of the underlying populations~\citep{LIGOScientific:2018jsj,Kimball:2020opk,Fishbach:2020ryj}. 
There are a diverse range of formation channels, and the true astrophysical population is likely to be a mixture drawn from a combination of channels. 
For example, BBH systems that merge within a Hubble time may form via the evolution of isolated field binaries that proceed through a common-envelope phase~\citep{Dominik:2012kk,Belczynski:2016obo,Stevenson:2017tfq,Eldridge:2018nop,Zevin:2020gma}, stable mass transfer~\citep{Heuvel:2017sfe,Klencki:2018fbx} or chemically homogeneous evolution~\citep{Mandel:2015qlu,Marchant:2016wow}; through dynamical encounters in dense stellar environments such as globular clusters~\citep{Downing:2009ag,Rodriguez:2016kxx,Askar:2016jwt,Rodriguez:2020viw}, young star clusters~\citep{PortegiesZwart:1999nm,Banerjee:2016ths,Rastello:2018elx,DiCarlo:2019pmf,Kumamoto:2020wqr} or nuclear clusters~\citep{Miller:2008yw,Antonini:2016gqe,Hoang:2017fvh,Sedda:2018znc,Zhang:2019puc}, which facilitate strong binary--single~\citep{Samsing:2017xmd,Arca-Sedda:2018qgq,DOrazio:2018jnv} and binary--binary~\citep{Arca-Sedda:2018qgq,Zevin:2018kzq} interactions to form hardened BBHs, or via the secular evolution of hierarchical systems undergoing Kozai--Lidov oscillations~\citep{Silsbee:2016djf,Rodriguez:2018jqu,Sedda:2020jvg}. 
Increasing the number of observed sources would help in identifying typical signatures of different channels~\citep{Mandel:2009nx,Stevenson:2017dlk,Talbot:2017yur,Zevin:2017evb,Sedda:2018nxm,Kalogera:2019sui,Sedda:2020vwo,Farmer:2020xne}. 
Each formation mechanism has its own associated uncertainties, inherent in our incomplete understanding of the underlying physics---this is exactly the physics we can come to understand through precision GW observations.

Current generation ground-based detectors will be able to detect stellar-mass BBHs out to redshifts of $z \sim 1$--$2$; next generation detectors like the Einstein Telescope~\citep{Sathyaprakash:2012jk} or Cosmic Explorer~\citep{Evans:2016mbw} will see out to $z \sim 20$, enabling them to chart the evolution of the binary population across the history of the Universe~\citep{Kalogera:2019sui}. 
To match the cosmological reach of upcoming ground-based detectors, and to stand a chance of making a non-negligible number of multiband detections, it is essential to have a detector with an enhanced sensitivity in the ${0.01}$--${1~\mathrm{{Hz}}}$ range. 
\textit{LISA} would only observe a small number of nearby ($z < 0.1$) systems~\citep{Moore:2019pke,Jani:2019ffg,Liu:2020eko}. 
This limited detection range means (i) that \textit{LISA} cannot probe evolution over cosmic time, it cannot even match the detection range of current ground-based detectors, and (ii) that multiband detections will be rare, making population inferences difficult if not impossible. 
An improved decihertz sensitivity will allow a more thorough characterization of the full range of formation channels which produce BBHs. 
DOs can potentially match, and even exceed, the range of next-generation ground-based GW observatories. 
Detection horizons for a range of binaries are shown in Figure~\ref{Fig2} and Table~\ref{tab:Horizons}.
BBHs could be observed from cosmic dawn to the early Universe ensuring that (i) by combining ground-based and decihertz observations we will have a \emph{complete multiband census} of the BBH population, and (ii) independent of ground-based detectors, DOs can perform a synoptic survey measuring not only the properties of the BBH population, but how the properties change with redshift. 
Of order of $10^3$ observations are required to place percent-level constraints on the currently uncertain parameters describing binary evolution~\citep{Barrett:2017fcw}.
Current ground-based detectors may achieve this after a few years at design sensitivity~\citep{Aasi:2013wya,LIGOScientific:2018jsj}; next-generation ground-based detectors and DOs can achieve the same in a couple of years for each redshift interval of width $0.1$~\citep{Kalogera:2019sui}. 
The large number of BBHs across a wide range of masses observable with DOs presents a remarkable opportunity to precisely determine the physics of binary evolution.

Less massive binaries, such as BNSs, are not observable to as great a distance; however, DOs still present the opportunity to observe these sources. 
Depending upon the DO design, the detection range varies significantly (see Table~\ref{tab:Horizons}).  
In more modest scenarios, the range is comparable to the current generation of ground-based detectors, and in the more ambitious cases we can find sources back to $z \sim 10$, when the Universe was only $500~\mathrm{Myr}$ old. 
Therefore, in the more optimistic cases, we can perform a census of the BNS, and NSBH binary, populations across cosmic time. 
Crucially, observations can extend beyond the peak in star formation rate at $z \sim 2$~\citep{Madau:2014bja}, enabling reconstruction of the delay time distribution between star formation and eventual merger~\citep{Belczynski:2018ptv,Safarzadeh:2019pis}. 
In the more pessimistic cases, we cannot perform the same population studies; however, decihertz observations still provide valuable warnings of a merger. 
This is especially beneficial for close-by binaries~\citep[e.g.,][]{Sedda:2020wzl}, as these are the most promising candidates for having observable multimessenger counterparts (GW170817 was at redshift $0.01$~\citep{TheLIGOScientific:2017qsa,LIGOScientific:2018mvr}). 
The decihertz observations provide notice of when a binary will merge, enabling telescopes to be positioned ready to observe the merger of the binary. 

The decihertz frequency range is ideal to study the detailed properties of the cosmic population of merging compact binaries. 
Ground-based detectors (even the next-generation instruments) are sensitive to the final phase of the inspiral, merger and ringdown, while space-based observatories probe the earlier inspiral phase. 
The two observations provide complementary information about the source, enabling improved parameter measurements. 
The symmetric mass ratio $\eta=(m_1 m_2)/(m_1 + m_2)^2$, where $m_1$ is the primary mass and $m_2$ is the secondary mass, is often not measured precisely through observations in a single frequency band~\citep{Blanchet:2013haa}. 
For BBHs, ground-based observatories measure more precisely the total mass $M = m_1 + m_2$ from the merger--ringdown phase, while space-based observatories measure more precisely the chirp mass $\mathcal{M} = \eta^{3/5}M$ from the inspiral phase~\citep{Isoyama:2018rjb,Cutler:2019krq,Marsat:2020rtl}. 
Combining the two can yield a much improved measurement of $\eta$~\cite{Vitale:2016rfr,Liu:2020nwz}. 
The mass ratio is correlated with the effective spin parameter $\chi_\mathrm{eff}$ (a mass-weighted sum of the individual component spins)~\citep{Damour:2001tu,Abbott:2017vtc}, and so it is possible to obtain more precise spin measurements. 
Further information about the spins can be obtained from observing precessional dynamics~\citep{Apostolatos:1994mx,Chatziioannou:2014coa,TheLIGOScientific:2016wfe}. 
Spin precession effects are easier to observe over long inspirals. 
This ensemble of precision information is useful both when looking at the overall population characteristics and when considering a single binary observed in both bands. 

\begin{figure}[ht!]
\centering
 \includegraphics[scale=0.3]{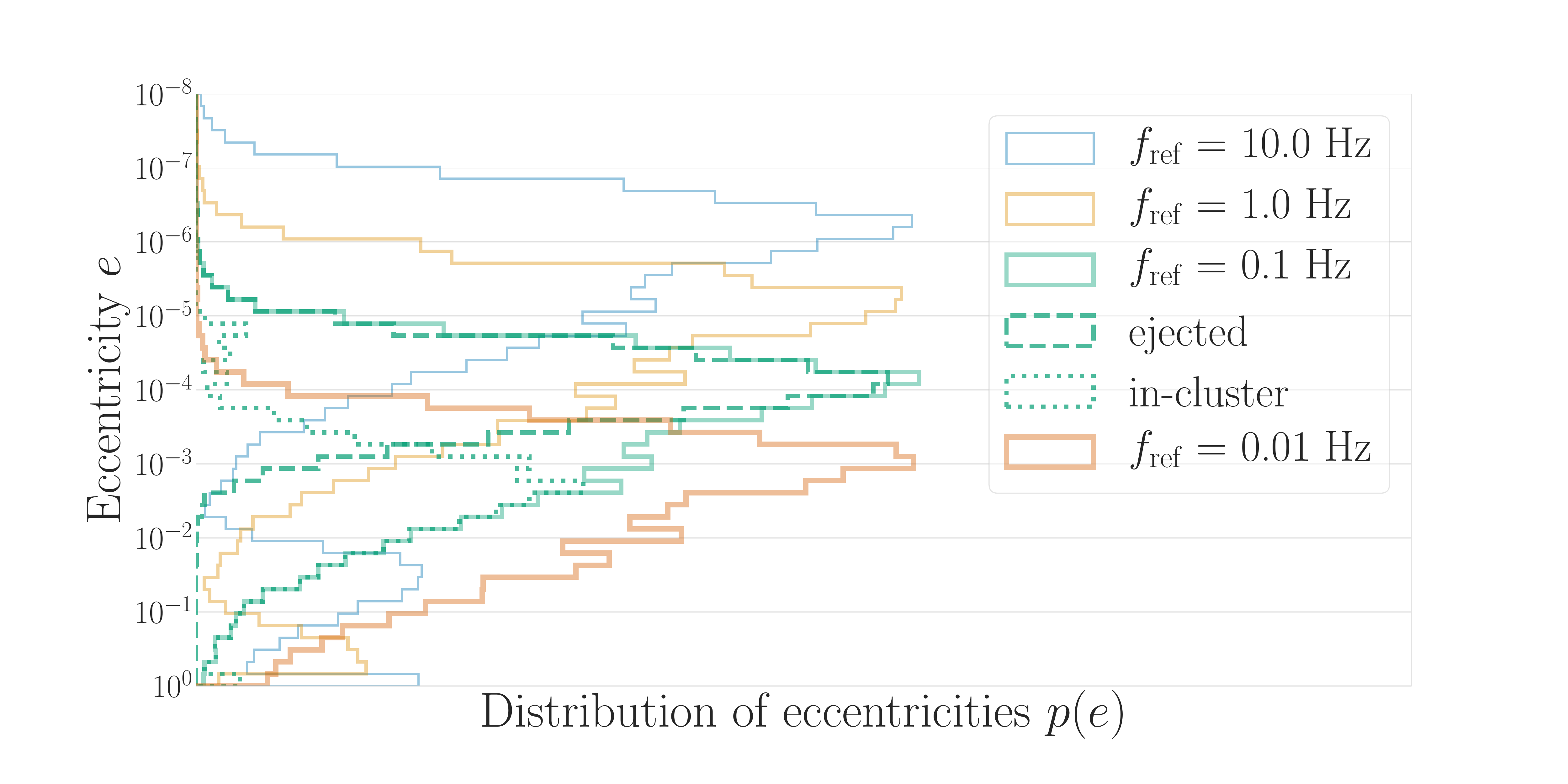} 
\caption{Eccentricity distributions of binary black holes formed in globular clusters when observed at different gravitational-wave frequencies. 
Binary binary black hole mergers in globular clusters form $3$ distinct populations: binaries that are ejected from the cluster due to a strong encounter, binaries that leave a strong encounter in a hardened state and merge before the next encounter, and binaries that merge during the strong encounter itself~\citep{Samsing:2017rat,Rodriguez:2018pss,Zevin:2018kzq}. 
The first $2$ populations make up the broad peak at lower eccentricity, and the third results in the distribution at $e \lesssim 10^{-2}$ seen for the gravitational-wave frequencies of $f_\mathrm{ref} = 10.0~\mathrm{Hz}$ and $f_\mathrm{ref} = 1.0~\mathrm{Hz}$. 
At lower frequencies it is easier to distinguish between the ejected and in-cluster merger populations~\citep{DOrazio:2018jnv,Samsing:2018nxk}; the dashed and dotted green lines differentiate the ejected and in-cluster populations, respectively, at $f_\mathrm{ref} = 0.1~\mathrm{Hz}$. 
The peak near $e \sim 1$ in the $f_\mathrm{ref} = 10.0~\mathrm{Hz}$ histogram is populated by systems that form in-band and merge on the timescale of days--years. 
}
\label{fig:ecc}
\end{figure}

In addition to masses, spins and merger rates, eccentricity is a key indicator of binary formation mechanisms.
Eccentricity is informative because binaries formed in isolation exhibit relatively lower eccentricities compared to those which evolved in dynamically active environments, such as dense stellar clusters or triples~\citep{Belczynski:2016obo,Rodriguez:2016kxx, Antonini:2016gqe,Antonini:2017ash,Samsing:2017xmd,Gondan:2017wzd,Arca-Sedda:2018qgq,Zhang:2019puc}. 
In addition to distinguishing field binaries from those formed in clusters or triple systems, eccentricity measurements can differentiate between sub-populations that are synthesized in dynamical environments, such as BBHs that are ejected from their host clusters and those that merge within the cluster environment. 
Dynamically-formed BBHs also have different characteristic eccentricities depending on whether they merge inside or outside of their host cluster, as illustrated in Figure~\ref{fig:ecc}.
However, eccentricity is hard to detect at higher frequencies due to the circularizing effects of gravitational radiation~\citep{Peters:1964zz}.
The residual eccentricity in the frequency range accessible by ground-based detectors is expected to be too small to be detectable, except in the rare cases when systems form with extreme eccentricities and rapid inspiral timescales~\citep{TheLIGOScientific:2016htt,Samsing:2017rat,Rodriguez:2018pss,Arca-Sedda:2018qgq,Zevin:2018kzq}. 
At millihertz frequencies, the eccentricity is much larger; \textit{LISA} would be able to distinguish between isolated and dynamical BBH formation channels through eccentricity measurements~\citep{Nishizawa:2016jji,Breivik:2016ddj,Nishizawa:2016eza,Canuel:2017rrp,Kremer:2018tzm,Randall:2019znp}, but only for a few nearby systems. 
In some cases, BBHs formed with the highest eccentricities will radiate GWs with frequencies that are too high for \textit{LISA} to observe~\citep{Randall:2018lnh,DOrazio:2018jnv,Kremer:2018tzm,Arca-Sedda:2018qgq,Kremer:2018cir,Zevin:2018kzq,Chen:2017gfm}. 
DOs are therefore well suited for eccentricity measurements. 
Furthermore, decihertz frequencies are ideal for observing BBH mergers in clusters that result from gravitational Bremsstrahlung---relativistic single--single encounters that dissipate enough energy during a close passage to become bound and merge~\citep{Samsing:2019}.
Decihertz observations will provide measurements of eccentricity, and hence insights into binary formation, unavailable in other GW bands.

\subsection{Localizing binaries \& identifying their host environments}

Longer duration observations provided by a space-based detector can provide improved sky localization compared to ground-based instruments. 
The motion of the constellation over its orbit provides information on the sky location~\citep{Takahashi:2003wm,Wen:2010cr,Mandel:2017pzd}. 
As well as providing information on spins, precession measurable in the inspiral also helps break the distance--inclination degeneracy~\citep{Vecchio:2003tn}. 
For nearby events with high signal-to-noise ratio, combining distance and sky localization will allow the unambiguous identification of host galaxies~\citep{DelPozzo:2017kme}.
This will provide otherwise inaccessible information about the connection between binary formation and the relevant galaxy properties, as well as a tool for constructing the local distance ladder through standard siren measurements. 
Standard siren measurements can be made with or without an electromagnetic counterpart~\citep{Schutz:1986gp,MacLeod:2007jd}, although results without a counterpart are in general much less precise~\citep{Chen:2017rfc,Abbott:2019yzh}.
Stellar-mass BBHs fall in this latter category~\citep{Kyutoku:2016zxn,DelPozzo:2017kme} and thus, although observed at larger redshift, are not expected to yield cosmological constraints comparable to BNSs, for which there are prospects for multimessenger counterparts~\citep{GBM:2017lvd}. 
From the early inspiral, both the localization and the merger time are known ahead of the merger.
Days or even years of warning (depending upon the properties of the binary and the orbital frequency observed) would enable electromagnetic observatories such as the Vera Rubin Observatory, \textit{WFIRST} and SKA to be positioned to observe at the time of the merger, providing the best possible coverage of the source. 
If each BNS can be assigned a redshift measurement from its host galaxy, then in the most optimistic scenarios we can construct a Hubble diagram with more than $\sim10^5$ events out to $z\sim3$~\citep{Cutler:2009qv,Nishizawa:2010xx}. 
With these measurements we can obtain subpercent constraints on the Hubble constant, and probe the equation of state of dark energy at the $10\%$ level or better~\citep{Cutler:2009qv}.
In addition to standard siren measurements, another cosmological application of the localization of binaries is mapping the large scale structure of the Universe. 
We will thus be able to probe the anisotropic structure of the Universe, independent of electromagnetic tracers, which enables us to map the cosmic matter distribution.
While multimessenger observations of BBHs are only possible if there is sufficient surrounding material, BNSs and NSBHs offer a rich source of emission. 
Capturing the early emission from the kilonova is particularly valuable in understanding the nature of NS matter, and hence forewarning of the merger is extremely valuable~\citep{Arcavi:2018mzm}.
These observations can tell us about the material properties of nuclear density matter~\citep{Abbott:2018exr,Montana:2018bkb,Most:2018hfd,Coughlin:2018fis,Margalit:2019dpi}, the production of heavy elements~\citep{Abbott:2017wuw,Chornock:2017sdf,Tanvir:2017pws,Wanajo:2018wra,Siegel:2018zxq}, and provide a unique laboratory for testing gravity~\citep{Monitor:2017mdv,Abbott:2018lct,Belgacem:2018lbp,Belgacem:2019pkk}. 
Even higher angular resolution (subarcminute) could be possible with a constellations of DOs~\citep{Baker:2019ync}.

A new insight made available from longer decihertz observations could be the measurement of a binary's centre-of-mass acceleration.
The peculiar acceleration of the center of mass of a compact binary leaves an imprint on the GW signal~\citep{Bonvin:2016qxr}. 
This can produce a detectable drift in the GW phase if the acceleration is large enough~\citep{Wong:2019hsq,Tamanini:2019usx}.
This drift is easier to find in longer signals (which ideally can be observed for the whole duration of the mission), and hence is a more promising target for space-based observatories than ground-based observatories, and for BNSs (and NSBHs) rather than BBHs in the decihertz frequency band~\citep{Seto:2001qf,Nishizawa:2011eq}.
Measuring the peculiar acceleration provides unique insight into the astrophysical environment of the source: for example it could reveal the possible presence of a third circumbinary object~\citep{Robson:2018svj,Wong:2019hsq,Tamanini:2019awb,Danielski:2019rvt} or indicate if the binary was embedded in a globular cluster~\citep{Inayoshi:2017hgw,Randall:2018lnh}.

\subsection{Observing double white dwarf mergers}
\label{Sec:WD}

Having access to decihertz frequencies would also enable multimessenger observations of WD binary mergers.
WDs are not accessible to ground-based detectors, as they merge at too low a frequency~\citep{Littenberg:2019grv}. 
A double WD binary with secondary mass $m_2$ ends its inspiral at $\sim 0.06 (m_2/\Ms)~\mathrm{Hz}$. 
Thus while WD binaries with low secondary masses (e.g., AM CVn-like progenitors~\citep{Nelemans:2004pp}) will either undergo stable or unstable mass transfer in the \textit{LISA} band~\citep{Marsh:2003rd,Gokhale:2006yn,Kremer:2017xrg}, most double WD binaries, including Type Ia progenitor candidates, reach the decihertz regime. 
By measuring this high-frequency population of double WDs, we can directly test the efficacy of the currently uncertain double-degenerate Type Ia supernovae (SNe) channel. 
A DO will be able to resolve this uncertainty as well as constrain the formation and evolution of Type Ia double WD progenitors~\citep{Hillebrandt:2013gna,Maoz:2013hna}. 
The best prospects for WD observations are achieved for DOs with detection horizons of $\gtrsim 20~\mathrm{Mpc}$, which enables detection of sources in the Virgo Cluster~\citep{Mandel:2017pzd}.
Based on observations of the Galactic double WD population~\citep{Badenes:2012ak}, $\sim 150$ double WD Type Ia progenitors would be visible in the Virgo Cluster, while population models suggest rates of $\sim 7$--$12$ times lower~\citep{Toonen:2012jj}. 
In the case of double WD binaries below the mass threshold to be Type Ia progenitors, the stability of mass transfer between WDs is still not well understood. 
It has been suggested that all mass transferring double WDs undergo novae which lead to a common-envelope like evolution leading to mergers for most double WDs~\citep{Shen:2015}. 
Observations in the decihertz frequency range will help resolve this mystery. 
Finally, potentially small numbers of WD binaries observed with high frequencies in the Milky Way will have enormous GW signal-to-noise ratios. 
These highly characterized systems will allow measurement of deviations from pure gravitational evolution, probing dissipative effects like tides~\citep{Benacquista:2011gh,Piro:2011qe,Fuller:2011jb,Fuller:2012dt,McNeill:2019rct,Kuns:2019upi}. 
Therefore, DOs will provide the opportunity to collect new observational insights into the WD population.

\subsection{Deciphering the physics of core-collapse supernovae}

Theoretically, SN explosions are powered either by explosive nuclear burning or by gravity driven core-collapse (CCSNe). 
The latter, among the most powerful explosions in the Universe, represent the culminating evolutionary stage of stars more massive than $\sim 8 \Ms$. 
Most of the GW energy released during a CCSN has frequencies in the range $\sim 10^2$--$10^3~\mathrm{Hz}$ and is due collectively to rapid $g$/$f$-mode oscillations in the newborn proto-NS (PNS)~\citep{Murphy:2009dx,Morozova:2018glm,Radice:2018usf}, turbulent convective motions, rotation, and (at times) a coherent standing-accretion-shock instability (SASI). 
However, during the explosion, the high-frequency emission is supplemented by two strong, low-frequency signals. 
The first is generated by asymmetric ejection of material over $0.1$--$1~\mathrm{Hz}$~\citep{Burrows:1995bb,Murphy:2009dx}. 
This emission is expected also for SNe developing in tight binaries~\citep[e.g.,][]{Holgado:2019vju}.  
The second low-frequency signal comes from asymmetric neutrino emission~\citep{Epstein:1978dv,Turner:1978jj,Burrows:1995bb,Muller:2011yi}. 
Outgoing neutrino shells produce GWs at frequencies of $\sim 0.1$--$10~\mathrm{Hz}$, thus overlapping with GWs from matter ejecta. 
Measuring this low-frequency emission with DOs would complement other observations to help us understand CCSNe.

\begin{figure}
\centering
\includegraphics[width=\textwidth]{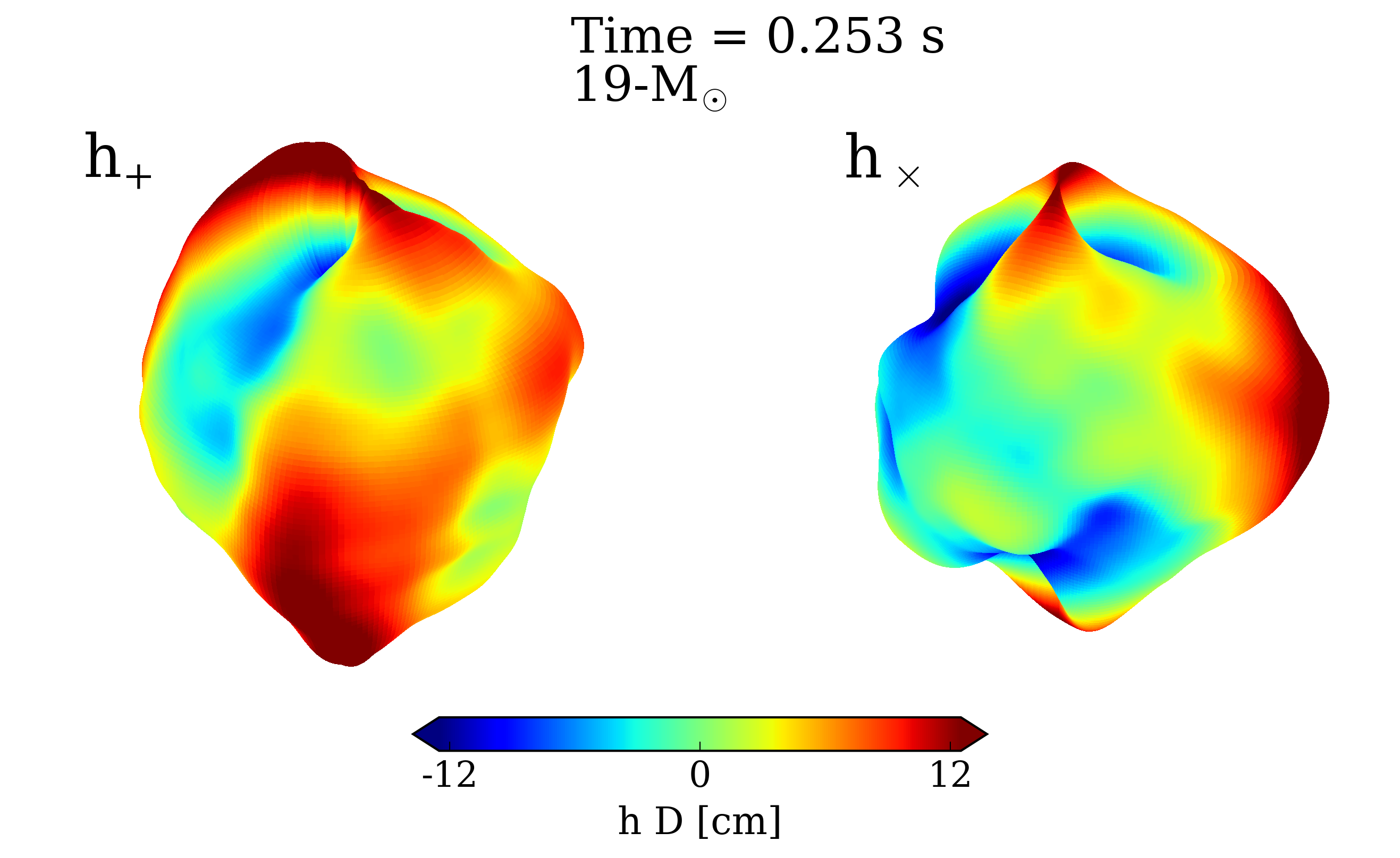}
\caption{Three-dimensional map illustrating the gravitational-wave strain $h$ (multiplied by distance $D$) generated by neutrino emission anisotropies $253~\mathrm{ms}$ after the SN bounce, assuming a stellar progenitor with mass $M = 19 \Ms$~\citep{Vartanyan:2020nmt}. 
The signal is shown for both $h_+$ (\emph{left}) and $h_\times$ (\emph{right}) polarization, and as a function of the viewing angle. Hotter colors (yellow to red; convex surfaces), indicate positive strains, whereas cooler colors (blue to yellow; concave surfaces) indicate negative strains.}
\label{fig:GWnu}
\end{figure}

Figure~\ref{fig:GWnu} shows the $h_+$ and $h_\times$ polarization of the GW signal generated by anisotropic neutrino emission at a representative time after the bounce of the core of a $M = 19\Ms$ star undergoing CCSN, as calculated using the {\textsc{Fornax}} supernova simulation code~\citep{Burrows:2019zce}. 
The two polarizations are different at any given time due to the stochasticity of the angular distributions of the emitted neutrinos. 
In the decihertz frequency range, anisotropic neutrino emission is more than one order of magnitude larger than GW signal emitted by anisotropic matter expulsion~\citep[e.g.,][]{Vartanyan:2019ssu,Vartanyan:2020nmt}.
At low frequencies, the metric strain due to anisotropic neutrino emissions dominates all other components, and also leaves a net metric displacement, similar to that associated with classical Christodoulou memory~\citep{Christodoulou:1991cr,Thorne:1992sdb}. 

CCSNe close enough to be detected are expected to be rare. 
The rate of CCSNe inferred for the Milky Way is $\Gamma_\mathrm{CCSNe} \sim 2$ events per century~\citep[e.g.,][]{Diehl:2006cf,Li:2010kd}. 
A back-of-the-envelope calculation of the amplitude of the GW signal emitted through this emission, accounting for the neutrino luminosity ($L_\nu \sim 10^{52}$ erg s$^{-1}$) and the duration of the neutrino burst ($\Delta t\sim 1~\mathrm{s}$), returns a characteristic strain $h_\mathrm{c} \sim 5\times 10^{-21}$ for sources at a distance $d = 10~\mathrm{kpc}$~\citep{Mueller:2003fs,Maggiore:2018sht}. 
Assuming an emission frequency of $\sim 1~\mathrm{Hz}$, the GW signal associated with Galactic CCSNe thus falls above the sensitivity curve of \textit{DECIGO}, DO-Optimal and DO-Conservative. 
Sources might be observable with DO-Optimal up to $\sim 1~\mathrm{Mpc}$ or with \textit{DECIGO} up to $\sim 10~\mathrm{Mpc}$, provided that the signal peak frequency is below a few hertz. 
Though probably hard to catch given the low rate, observing GWs from CCSNe with DOs, in combination with their higher frequency GW emissions, would provide insights into (i) the neutrino emission and matter ejection occurring during a CCSN, and (ii) the physics of SN explosions.

%%%%%

\section{Uncovering the formation and evolution of intermediate-mass black holes}
\label{Sec:IMBH}

\subsection{Intermediate-mass black holes in star clusters: Intermediate mass-ratio inspirals}

IMBHs are an elusive class of BHs with masses in the range $10^2$--$10^5 \Ms$, which are expected to bridge stellar-mass and massive BHs.  
IMBH observations would provide insights about stellar evolution and dynamics, while excluding the existence of IMBHs across a portion of the mass spectrum would prove that massive BHs must form from heavier seeds, and not grow from stellar-mass BHs. 
One of the proposed formation scenarios for IMBHs formation is via repeated mergers of stars and compact stellar remnants in dense star clusters, taking place either on short timescales ($<1~\mathrm{Gyr}$)~\citep{PortegiesZwart:1999nm,Giersz:2015}, or throughout the whole host cluster lifetime ($>1~\mathrm{Gyr}$)~\citep{Giersz:2015,Sedda:2019rfd}. 
The relatively high densities required to trigger IMBH seeding and growth makes globular clusters ideal places to look for IMBH signatures. 
The conclusive identification of an IMBH in a globular cluster has not yet been made~\citep{Mezcua:2017npy}, owing to the small effect that an IMBH has on the surrounding stars. 
For instance, the foremost kinematical measurements available are not yet sufficiently detailed to resolve the IMBH influence radius, leading sometimes to controversial results~\citep{Lanzoni:2013zfa}.
Unfortunately, the robustness of the results does not necessarily improve using other measurements techniques, like detection of stellar disruption events~\citep{Lin:2018dev}, or millisecond pulsar timing~\citep{Colpi:2003wb}. 
GWs provide a new, conclusive means of IMBH detection.

As the heaviest objects in a star cluster, IMBHs have a high probability to form a binary with a compact stellar remnant, like a WD, NS, or stellar-mass BH~\citep{Konstantinidis:2011ie,MacLeod:2015bpa,Sedda:2019rfd}. 
Observing the coalescence of an IMBH and a compact remnant would provide us with a wealth of information. 
First, such an observation indicates that IMBHs do exist in that portion of the the mass spectrum. 
Second, the nature of the IMBH companion would allow us to discover their nursing environments. 
A WD companion would most likely imply the absence of stellar-mass BHs in the cluster~\citep{Sedda:2019rfd}; a stellar-mass BH companion would imply that stellar-mass BHs can co-exist with an IMBH, and a NS companion would allow us to place constraints on NS natal kicks. 
Mergers involving a WD or a NS can lead to bright electromagnetic counterparts associated to the stellar tidal disruption~\citep{Chen:2018foj,Eracleous:2019bal}; these multimessenger observations provide insights into WD or NS structure.

GW astronomy will offer a unique opportunity to uncover the details of the IMBH population. 
The typical mass ratio between a stellar companion and an IMBH lies in the range $q = m_2/m_1 \sim 10^{-2}$--$10^{-4}$, thus IMBH-stellar remnant binaries are referred to as IMRIs~\citep{AmaroSeoane:2007aw,Brown:2006pj,Rodriguez:2011aa}. 
More extreme mass ratios, typical of binaries containing a massive BH and a stellar companion, are called extreme mass-ratio inspirals (EMRIs)~\citep{AmaroSeoane:2012tx,Amaro-Seoane:2014ela,Berry:2019wgg}. 
While the physical processes behind EMRI formation are relatively well known, being mostly due to two-body relaxation~\citep{AmaroSeoane:2012tx,Alexander:2017rvg} and to the massive BH spin~\citep{AmaroSeoane:2012cr,Babak:2017tow}, the picture is less constrained in the case of IMRIs. 
The loss-cone theory, which is at the basis of EMRI evolution~\citep{Heggie:2003,Spitzer:1987}, is not yet fully understood for IMBHs; this problem is more complicated because the IMBH is not fixed in the centre of the host cluster. 
Numerical models suggest that IMRIs forming deep inside star clusters are characterized by large eccentricities at formation (from $e=0.999$) and small semi-major axis (below $a \sim 10^{-5}~\mathrm{pc}$)~\citep{Konstantinidis:2011ie,Leigh:2014oda,Hong:2015aba,MacLeod:2015bpa,Haster:2016ewz}. 
As GW emitters, IMRIs can be jointly detected with ground-based observatories and space-based observatories~\citep{AmaroSeoane:2009ui,Jani:2019ffg}. 

While IMBH formation scenarios will remain uncertain until we obtain observations, it is possible to place constraints on IMRI merger rates by making minimal assumptions on the star cluster formation rate per unit mass $\rho_\mathrm{SFR}$, the fraction of clusters hosting an IMBH $p_\mathrm{IMBH}$, and the fraction of IMBHs developing an IMRI $p_\mathrm{IMRI}$. 
For simplicity, we assume that $\rho_\mathrm{SFR}$ is nearly constant across the redshift range $2$--$8$~\citep{Katz:2012nd} and that the fraction of IMBHs having a stellar-mass BH companion is $f_\mathrm{com} \sim 0.01$~\citep{Sedda:2019rfd}.
During a $4~\mathrm{yr}$ long mission, the \textit{LISA} will enable us to observe a typical IMRI with component masses $10^3 \Ms+30 \Ms$ out to a redshift $z\simeq 0.2$ with SNR $15$. 
A rough but informative estimate of the corresponding IMRI detection rate can be written as the product of the number of times an IMBH forms an IMRI $n_\mathrm{rep}$, the fraction of clusters hosting an IMBH $p_\mathrm{IMBH}$, the fraction of clusters in which an IMRI can form $p_\mathrm{IMBH}$, the fraction of a stellar ensemble comprised of a given type of compact objects $f_\mathrm{com}$, the star cluster formation rate per unit mass $\rho_\mathrm{SFR}$, the average cluster mass $M_\mathrm{GC}$, and the sensitive volume at a given redshift $V$ \citep[e.g.,][]{Chen:2017wpg}. 
Assuming typical values we have:
\begin{eqnarray}
    \Gamma_\mathit{LISA} \simeq & 0.012 n_\mathrm{rep}\left(\frac{p_\mathrm{IMBH}}{0.2}\right)
    \left(\frac{p_\mathrm{IMRI}}{0.5}\right)
    \left(\frac{f_\mathrm{com}}{0.01}\right) \times \nonumber \\
    & \times \left(\frac{\rho_\mathrm{SFR}}{0.005 \Ms~\mathrm{yr}^{-1}\,\mathrm{Mpc}^{-3}} \right)\left(\frac{M_\mathrm{GC}}{10^6\, \Ms}\right)^{-1}\left(\frac{V}{2.5~\mathrm{Gpc^3}}\right)~\mathrm{yr}^{-1}.
\end{eqnarray}
A DO characterized by a larger detection horizon would boost these rates: in the case of IMBH--BH systems \textit{ALIA} enhances the prospect of detection by a factor $\Gamma_\mathit{ALIA} \sim 1380 \Gamma_\mathit{LISA}$, similarly for \textit{DECIGO}, while our DO-Conservative and DO-Optimal designs leads to $\Gamma_\mathrm{DO} \sim 1115$--$1217 \Gamma_\mathit{LISA}$, respectively. 
To infer the rate for IMRIs containing a WD or a NS, we must take into account that the horizon redshift will change compared to IMRIs containing a stellar BH, and that the probability for an IMBH to have a WD or NS companion will vary as well, being $f_\mathrm{com} \sim 0.16$ in the case of WDs and $f_\mathrm{com} \sim 0.002$ for NSs~\citep{Sedda:2019rfd}. 
The simple calculation presented here neglects many effects that need detailed and careful treatment, like globular clusters might not form beyond redshift $z>2$--$8$, and that the growth of an IMBH and the establishment of an IMRI take place over a cluster relaxation time. 
Nonetheless, even in the case in which IMBHs formation is limited to the redshift range $z=2$--$8$, and considering only IMRIs containing a stellar BH, the boost factor gained with any of the DOs discussed here would be $\sim 165$--$606$ times larger than for \textit{LISA}, making this class of observatories a crucial element to unveil the physics and dynamics behind IMBHs. 
Table~\ref{tab3} summarizes IMRI merger rates for different detectors, and the number of events per year involving a WD, NS, or stellar-mass BH. 
As a consequence of the frequency range of IMRI signals, a DO would enhance the prospects of detection to tens of events per year, and allow us to observe potential IMRI populations up to high redshift (up to $z\sim 300$ with \textit{ALIA}, which is sufficient to capture essentially all IMBH--BH IMRIs in the Universe).

\begin{table*}[]
 \begin{center} 
  \caption{Prospective intermediate mass-ratio inspiral detection rates with different gravitational-wave observatories.
  Horizon redshift $z_\mathrm{H}$ and event rates $\Gamma$ are shown for inspirals of white dwarfs, neutron stars and stellar-mass black holes. The rates are calculated assuming that the probability of an intermediate-mass black hole forming in a cluster is $p_\mathrm{IMBH} = 0.2$, that the probability of an intermediate-mass black hole forming an intermediate mass-ratio inspiral is $p_\mathrm{IMRI} = 0.5$, and that each of these intermediate-mass black hole forms an intermediate mass-ratio inspiral $n_\mathrm{rep} = 1$ times in the same cluster. 
  We assume that the probability for an IMBH to have a (black hole, neutron star, white dwarf) companion is ($1\%, 0.25\%, 16\%$) respectively~\citep{Sedda:2019rfd}.}
  \label{tab3}
  \begin{tabular}{lcccccc}
         \br
         Detector & \multicolumn{2}{c}{Black holes} & \multicolumn{2}{c}{Neutron stars} & \multicolumn{2}{c}{White dwarfs} \\ 
                  &  $z_\mathrm{H}$  & $\Gamma/\mathrm{yr}^{-1}$ & $z_\mathrm{H}$ & $\Gamma/\mathrm{yr}^{-1}$ & $z_\mathrm{H}$ & $\Gamma/\mathrm{yr}^{-1}$ \\
          \mr         
          \textit{LISA}       & 0.2 & 0.012 & 0.013 & $9.8 \times10^{-7}$ & 0.0052 & $0.0038$ \\
          DO-Conservative     & 46  & 13    & 1.7  & 0.4                 & 0.9    & 7.6  \\
          DO-Optimal          & 78  & 14    & 18  & 2.6                 & 7.8    & 110 \\
          \textit{DECIGO}     & 250 & 16    & 92    & 3.6                 & 70     & 220 \\
          \textit{ALIA}       & 300 & 16    & 3.9  & 1                 & 1.4    & 17  \\
          \br
    \end{tabular}
    \end{center}
\end{table*}

As with stellar-mass binaries, combining ground-based and space-borne observations can provide better constraints on the properties of the source. 
Space-based detectors can observe the inspiral and hence provide us with measurements of parameters such as the chirp mass~\citep{Isoyama:2018rjb,Cutler:2019krq,Toubiana:2020cqv}, while ground-based detectors will 
detect the merger and ringdown, and therefore measure other parameters such as the final mass and spin~\citep{Haster:2015cnn}. 
Therefore, the joint detection of IMRI GWs can allow us to break parameter degeneracies and place more stringent constraints on IMRI mergers. 
However, high-mass or high-redshift IMRIs will merge at frequencies which are too low for ground-based detectors, and DOs will become especially valuable as they will be able to measure the merger and ringdown.

Having access to IMRIs via DOs will provide a unique test-bed for assessing the accuracy of general relativistic waveforms. 
The orbiting companion object spends many orbits in the strong field regime close to the IMBH, enabling us to probe the structure of the spacetime.  
Computing the GWs in the mass ratio regime of IMRIs presents a challenge for current approaches to modelling the relativistic two-body problem. 
Although numerical relativity~\citep{Duez:2018jaf} can formally model these binaries, the mass ratio leads to extremely high spatial and temporal resolution requirements. 
Coupled with the need to model thousands of potentially highly eccentric orbits~\citep{Amaro-Seoane:2018gbb} this, currently, renders numerical-relativity simulations of these binaries impractical (though future algorithmic and technological developments may close this gap). 
One promising approach for modeling IMRIs is BH perturbation theory, which expands the Einstein equations in powers of the (inverse) mass ratio around the analytically known metric of the primary. 
Once this approach is taken to second order in the mass ratio the error in the waveform phase will be sufficiently small (depending on the coefficient of the unknown third-order term~\citep{Hinderer:2008dm}) to model IMRIs. 
With progress on second-order calculations well underway~\citep{Pound:2012nt,Gralla:2012db,Pound:2014xva,Wardell:2015ada,Miller:2016hjv,Pound:2019lzj} we expect that this technique will cover a large proportion of the IMRI parameter space. 
These approaches can also be coupled to post-Newtonian theory when the binary is widely separated~\citep{Blanchet:2013haa}, and to effective-one-body theory~\citep{Taracchini:2013rva,Bohe:2016gbl,Nagar:2018zoe}. 
The best IMRI waveform models will be created by combining the strengths of each of these approaches. 
Accessing the decihertz observational window will give us a unique chance to place strict constraints on the theory of IMRI GW emission and on the accuracy of our modelling techniques. 

In addition to the wealth of information that can be provided by GW observations alone, there is potential for IMRI systems to be multimessenger sources. 
A WD inspiralling around an IMBH with mass $\lesssim 10^5 \Ms$ will be tidally disrupted before it is swallowed by the IMBH. 
The slow WD inspiral will emit an IMRI GW signal accompanied by bright electromagnetic emission during the tidal disruption and accretion of the WD. 
The flare associated to the tidal disruption can significantly exceed the Eddington luminosity of $\sim 10^{43}(M/10^5\Ms)~\mathrm{erg\,s}^{-1}$~\citep{Sesana:2008zc,Zalamea:2010mv}. 
The BH mass range needed to trigger this mechanism is typical of IMBHs predicted to be sitting in globular clusters or in the centres of dwarf galaxies. 
Due to the poorly known demographics, event rates for WD--IMBH mergers are currently highly speculative and model-dependent. 
Calculations based on numerical models suggest event rates as high as $\sim 100~\mathrm{yr}^{-1}\,\mathrm{Gpc}^{-3}$ in dwarf galaxies and $\sim 1~\mathrm{yr}^{-1}\,\mathrm{Gpc}^{-3}$ in globular clusters, assuming that all dwarf galaxies and globular clusters host an IMBH~\citep{MacLeod:2014mha,MacLeod:2015bpa}, but they can be much lower depending on the unknown occupation fraction. 
For typical IMBH ($10^3 \Ms$) and WD ($0.8 \Ms$) masses, a DO detector can observe these systems at redshift $z\lesssim 1.8$;  
heavier IMBHs ($\sim 10^4 \Ms$) can be detected at even higher redshift ($z\simeq 3.5$), thus {joint electromagnetic and GW observations can potentially lead to the discovery of thousands of tidal disruptions}.

Through observations of IMRIs, we probe the low-mass end of massive BH demographics. 
This has a dual impact on our knowledge of BHs. 
First, we will have access to the mass range which contains the secrets of massive BH seeds~\citep{Volonteri:2009vh}. 
Understanding how massive BHs form and grow is key to understanding how structure forms in the early Universe and how galaxies evolve. 
Second, we will have the unique possibility to observe stellar dynamics of dense star clusters in action. 
Furthermore, the observation of electromagnetic counterparts to tidal disruptions will provide us with: (i) crucial details on accretion physics (and likely super-Eddington flows)~\citep{Dai:2018jbr}, and its dependence on the mass and spin of the accretor (determined by the GW observations); (ii) another powerful class of standard sirens for cosmography, and (iii) precise localization of the event, making possible a definitive association of IMBHs with globular clusters or dwarf galaxies.

\subsection{Intermediate- and stellar-mass black holes in galactic nuclei}

Galactic nuclei and nuclear star clusters are expected to contain a dense population of stellar-mass BHs orbiting around a (super)massive BH~\citep{Morris:1993zz,MiraldaEscude:2000cg}. 
These stellar-mass BHs can either be formed in-situ~\citep{Antonini:2014spa,Generozov:2018niv}, or via deposit from orbit segregated star clusters~\citep{Antonini:2014spa,Sedda:2018znc,Fragione:2020rmf}. 
The enrichment in BHs typical of this class of stellar systems is due to a variety of factors. 
The large escape velocities increase the probability to retain BHs, whereas mass segregation naturally causes a drift of the BHs towards the innermost galactic regions~\citep{Bahcall:1976aa,AmaroSeoane:2004hx,Freitag:2006qf,Hopman:2006xn,Alexander:2008tq,Berry:2013ara,Panamarev:2018bwq}. 
The high densities regulate both BBHs formation~\citep{Lee:1994nq,Gondan:2017wzd,Rasskazov:2019gjw}, hardening~\citep{Miller:2008yw} and ionization~\citep{Hopman:2009gz,OLeary:2008myb,Hong:2015aba,Leigh:2017wff} via dynamical encounters. 
The presence of a massive central body, either a massive BH or an IMBH, sitting at the heart of the galactic centre can have a crucial impact on the evolution of the star cluster and its population of BBHs. 

\subsubsection{Gas-rich galactic nuclei}

A dense AGN gas disc around the central massive BH dynamically cools the orbits, setting up a preferential orbital plane~\citep{Syer:1991}. 
Hence, a fraction of the nuclear BH population must end up on circularized orbits within the AGN disc~\citep{McKernan:2012rf}. 
The gas disc torques embedded objects, and differential migration within the disc allows new BBHs to form at low relative velocities~\citep{McKernan:2014oxa,Secunda:2018kar}. 
Disc gas and tertiary encounters can harden BBHs to merger, yielding a population of overmassive BBH mergers in AGN discs detectable as GW merger events with ground-based GW detectors~\citep{McKernan:2014oxa,Bartos:2016dgn,Stone:2016wzz,McKernan:2017umu,Yang:2019okq}. 
At large separations these embedded BBHs will be detectable in the \textit{LISA} GW band, where the effects of AGN gas drag on the binary could be imprinted on the GW waveform~\citep{Derdzinski:2018qzv,McKernan:2017umu}. 
Depending on AGN disc structure, gas torques can cause embedded migrating objects to converge at traps, where high-mass IMBHs can be built up~\citep{Bellovary:2015ifg,Secunda:2018kar,McKernan:2019beu,Yang:2019cbr}. 
As the IMBH builds up at the trap via mergers with in-migrating BH, the resulting IMRIs will appear as GW emitters in the $0.01$--$1~\mathrm{Hz}$ frequency band. 
An electromagnetic counterpart simultaneous with the IMRI will be detectable in the ultraviolet/optical due to Hill-sphere contraction at merger~\citep{McKernan:2019hqs}. 
IMBHs with masses $\sim 10^2$--$10^4 \Ms$ can also be brought into the innermost galactic regions by disrupting star clusters formed close to the galactic centre~\citep{Ebisuzaki:2001qm,PortegiesZwart:2005zp,Mastrobuono-Battisti:2014uaa,Sedda:2018znc,Arca-Sedda:2017wea}. 
These delivered IMBHs can further segregate toward the (super)massive BH via dynamical friction and get trapped into the AGN disc, thus potentially pairing with BHs, forming IMRIs, and potentially allowing the IMBH mass buildup. 
A DO has the potential to uncover the population of IMRIs and BBHs forming in AGN discs, detecting them years--weeks prior to the merger. 
{The associated GW signal encodes information about the AGN environments, providing a unique link between the small scale of the IMRI or BBH merger, and the large scale characterizing the AGN and the host galactic nucleus}.

\subsubsection{Gas-poor galactic nuclei}

A galactic nucleus harboring a quiescent (super)massive BH can be a site of intense GW source formation. 
Close to the (super)massive BH, IMBHs can either form via multiple stellar collisions~\citep{Antonini:2016gqe}, or be transported by inspiralling star clusters~\citep{Sedda:2018znc}, opening the possibility of the formation of (super)massive BH--IMBHs multiplets~\citep{PortegiesZwart:2005zp,Mastrobuono-Battisti:2014uaa,Arca-Sedda:2017wea}. 
Alternatively, the nuclear BH can tidally capture BBHs onto tightly bound orbits~\citep{Addison:2015bpa,Chen:2018axp} and affect the BBH evolution via the Kozai--Lidov effect~\citep{Antonini:2012ad,Prodan:2014dla,Stephan:2016kwj,VanLandingham:2016ccd,Liu:2017hho,Petrovich:2017otm,Bradnick:2017pww,Sedda:2018znc,Hamers:2018hxv,Hoang:2017fvh,Arca-Sedda:2017wea,Fragione:2018yrb}. 
In this case, a BBH orbiting a nuclear BH undergoes a periodic increase of its eccentricity, which can reach values close to unity and, consequently, shorten the binary lifetime. 
Binaries forming this way have low probability to retain a residual eccentricity in the ground-based observational band, but would be easier to observe with a DO~\citep{Miller:2002vg,Sesana:2016ljz,Chen:2017gfm,Hoang:2019kye}. 
The corresponding signal is long-lasting, likely $\sim3$--$5~\mathrm{yr}$, comparable to the lifetime of the detector.  
Such a long period of observation will allow us to discern the subtle distortion of the GWs caused by the orbital motion of the BBHs around the (super)massive BHs or IMBHs~\citep{Inayoshi:2017hgw,Meiron:2016ipr,Robson:2018svj,Arca-Sedda:2017wea,Chamberlain:2018snj,Wong:2019hsq}, the tidal force of the (super)massive BHs or IMBHs~\citep{Meiron:2016ipr,Hoang:2019kye,Randall:2018lnh,Fang:2019hir}, or the hydrodynamical effects in any accretion discs~\citep{Chen:2019jde}.

DOs enable multi-band observations of those BBHs merging in the vicinity of IMBHs, potentially within $\lesssim 10^2$ Schwarzschild radii~\citep{Chen:2017xbi}. 
BBH--IMBH systems are a unique target for DO--ground-based multiband observation because they emit not only ${1}$--${10^2~\mathrm{{Hz}}}$ GWs from the coalescing binary, but also simultaneously ${0.01}$--${1~\mathrm{{Hz}}}$ GWs due to the orbital motion of the binary around the IMBH. 
Moreover, because the remnant of the coalescing binary recoils due to anisotropic GW radiation~\citep{Centrella:2010mx,Blanchet:2013haa}, the orbit of the post-merger BH around the IMBH changes slightly, which leaves an imprint in the low-frequency waveform. 
By observing such a feature in the $0.01$--$1~\mathrm{Hz}$ band we can conduct a series of precise experiments, such as measuring the mass and linear momentum loss via GWs, as well as constraining the graviton mass. 
The corresponding precision would be an order of magnitude better than the current limits~\citep{Han:2018hby}.
Thanks to the large detection horizon and, consequently, the high detection rate of BBHs, a detector working in the $0.01$--$1~\mathrm{Hz}$ band can potentially detect multiple such events each year, even though such events are expected to be rare relative to BBH mergers formed from other channels.

\section{New frontiers of fundamental physics}
\label{Sec:fun}

\subsection{Tests of general relativity}

GW astronomy provides a powerful new toolbox which enables testing the laws of gravity at new scales and in new regimes.
Compact binary coalescences allow in particular to probe the strong-field regime close to the merger where the spacetime is dynamical~\citep{Gair:2012nm,Will:2018bme,Yagi:2013du}, as well as the cosmological propagation of GWs~\citep{Ezquiaga:2018btd,Shao:2020shv}.

Although GR has withstood all stress-tests to-date~\citep{Will:2014kxa, Berti:2015itd, Yunes:2016jcc, TheLIGOScientific:2016src,Abbott:2018lct,LIGOScientific:2019fpa}, it is not a viable candidate for quantum gravity. 
Furthermore, the theory does not provide a natural scale to explain the value of the cosmological constant~\citep{Clifton:2011jh}. 
Solving these puzzles remains an important problem in theoretical physics, with a plethora of alternative theories providing corrections to GR in the weak-field, low-density regime of cosmology. 
Testing these models is a key driver of upcoming cosmology and astrophysics experiments~\citep{Ishak:2019aay,Amendola:2016saw}.

A generic property of extensions to GR is the inclusion of new propagating degrees of freedom. 
These are sometimes introduced purposefully as a feature of the gravity model (e.g., Jordan--Brans--Dicke theory introduces a scalar field that couples conformally to the Einstein--Hilbert action~\citep{Ballardini:2019tho}), and in other models they are induced as a consequence of modifications to the dynamics of the metric (e.g., $f(R)$-gravity and non-local gravity~\citep{Sotiriou:2008rp,Berry:2011pb,Belgacem:2017cqo}). 
These may lead to novel solutions such as hairy BHs in quadratic gravity~\citep{Kanti:1995vq,Sotiriou:2014pfa,Maselli:2015yva,Benkel:2016rlz} or in the presence of time varying scalar fields on cosmological scales. 
Modified gravity models involving vector fields, like Einstein-\ae{}ther theory~\citep{Jacobson:2008aj}, and additional tensor fields, like bigravity~\citep{Schmidt-May:2015vnx}, are also possible.
A binary composed of such hairy BHs will emit additional (scalar) dipole radiation that causes a phase shift in the GW signal~\citep{Seymour:2018bce,Yagi:2016jml}, and measuring these enables placing the most stringent observational bounds on quadratic gravity to-date~\citep{Yagi:2012gp,Okounkova:2017yby,Witek:2018dmd}.

{Observing GWs in the decihertz range enables tests of GR in a new regime}. 
For example, in an alternative theory of gravity possessing a scalar field with a nonzero mass, the scalar field around the compact object will be confined inside its Compton wavelength, decaying exponentially outside this characteristic distance~\citep{Ramazanoglu:2016kul,Yazadjiev:2016pcb,Staykov:2018hhc}. 
Therefore, the dipolar radiation during the inspiral will be suppressed until the orbital separation between the two inspiralling compact objects drops below the characteristic Compton wavelength. 
This implies that the binary dynamics will change in different stages of the signal~\citep{Alsing:2011er,Sagunski:2017nzb}, making necessary the observation of a large range of frequencies to detect these effects. 
Alternatively, in some theories of gravity, BHs develop nontrivial hair only for a certain range of parameters, while in the rest of the parameter space the solutions coincide with GR. 
This happens for theories that admit scalarization similar to the Gauss--Bonnet gravity, where the development of a nontrivial scalar field is triggered by the spacetime curvature~\citep{Doneva:2017bvd,Silva:2017uqg}, but can also be extended to other classes of alternative theories~\citep{Andreou:2019ikc}. 
Moreover, for such theories an effect called dynamical scalarization (or descalarization) can be observed: BH scalar hair can develop (or vanish) as the orbital separation between the two compact objects changes~\citep{Barausse:2012da,Shibata:2013pra,Sennett:2016rwa,Khalil:2019wyy}.
Thus, only by observing different classes of BBHs across different mass ranges can we check for the appearance or disappearance of nontrivial hair. 
{Increasing the range of the frequencies observed increases the probability to observe the transition region, and, if none is found, enables us to verify the applicability of GR in a space only accessible with a DO}.

Further GW tests of GR can be boosted with joint space- and ground-based detections~\citep{Sesana:2016ljz}. 
The improved source characterization of multiband events~\citep{Vitale:2016rfr,Jani:2019ffg,Liu:2020nwz} leads to precision tests of parametrized deviations of GR~\citep{Carson:2019rda,Gnocchi:2019jzp,Toubiana:2020vtf}. 
Consistency checks can be performed measuring parameters like the masses and spins independently from space-based and ground-based detectors~\citep{Hughes:2004vw,Ghosh:2016qgn,Tso:2018pdv}. 
Moreover, tracking the phase and amplitude of the GW in such extended frequency band (possibly four orders of magnitude) enables to severely bound modified dispersion relations and frequency dependent modulations of the strain. For instance, constraints on GW oscillations could be highly improved with respect to \textit{LISA} capabilities~\citep{Belgacem:2019pkk}. 
With a DO working in concert with next-generation ground-based detectors, there is the potential to perform many multiband tests of the nature of gravity.

To illustrate one test where observations across the decihertz band would answer a key question about gravitation, let us study the \emph{strong equivalence principle}, one of the central pillars of Einstein's GR. 
A typical consequence of violating the strong equivalence principle is the emission of dipole radiation from asymmetric binaries~\citep{Will:2018bme,Barausse:2016eii}. 
Fields non-minimally coupled to the Einstein--Hilbert term or to the matter Lagrangian may have non-trivial profiles around massive bodies. 
This does not usually affect BHs, due to no-hair theorems~\citep{Damour:1993hw}, hence we generally consider binaries with at least one non-BH component; however, for certain classes of gravity theories, dipole radiation is also possible for a BBH system, evading the no-hair theorems~\citep{Barausse:2015wia,Yagi:2015oca}.
The non-trivial field profile results in a violation of the strong equivalence principle, because the gravitational properties of a body now depend on its internal scalar profile. 
As a result, the matter stress-energy tensor is generally not conserved, allowing for the emission of dipole radiation.

An unusual feature of dipole radiation is that its relevance (relative to quadrupole radiation) decreases as the binary separation shrinks. 
Hence, it is crucial to observe the evolutionary phase of the binary at which it is close enough to have significant GW emission, but still far from rapid inspiral and merger. 
DOs sensitive at $0.01$--$1~\mathrm{Hz}$ would enable a significant number of binaries to be followed through this key phase. 
The dipole flux correction to the total GW flux $\dot{E}_\mathrm{GW}$ can be quantified using a parameter defined through~\citep{Barausse:2016eii}
 \begin{eqnarray}
 \dot{E}_\mathrm{GW} & = \dot{E}_\mathrm{GR}\left[1+B\left(\frac{GM}{rc^2}\right)^{-1}\right],
\end{eqnarray}
where $M$ is the total binary mass, $r$ is the orbital separation, and $\dot{E}_\mathrm{GR}$ is the flux predicted in GR. 
A combined DO and next-generation ground-based detector network could constrain $B<10^{-12}$--$10^{-10}$, which is orders of magnitude better than the results from ground-based detectors and \textit{LISA}~\cite{Liu:2020nwz}. 
These generalised constraints can then be mapped onto specific classes of theories within the modified gravity landscape, validating them or ruling them out.

This discussion of dipole radiation does not consider the issue of screening. 
For gravity theories that modify the large-scale cosmological regime, a mechanism is needed to ensure consistency with other weak-field regimes, such as the Solar System. 
Screening refers to a handful of mechanisms that are known to operate in some (but not all) modified gravity theories, that act to suppress their effects in regions of high density~\citep{Joyce:2014kja}. 
In screened theories, one generally expects that the entire interior or a galaxy is screened, and hence governed by the laws of GR. 
Currently known screening mechanisms are proven to work only in static gravitational regimes. 
One of the most common screening mechanisms, the Vainshtein mechanism~\citep{Vainshtein:1972sx,Babichev:2013usa}, is known not to operate fully in \emph{dynamical} gravitational situations~\citep{Jimenez:2015bwa}. 
{This makes GW tests of gravity not only possible, but also potentially the \emph{only} way to probe modified gravity effects inside galaxies}.

Multimessenger observations have already had a substantial impact on the landscape of modified gravity through constraints on the GW propagation speed obtained from the GW170817 event~\citep{Monitor:2017mdv,Abbott:2018lct} and its electromagnetic counterpart. 
Many popular gravity models prior to GW170817 predicted an anomalous GW speed $c_\mathrm{T}(z)$ at low redshifts. 
GW170817 alone implies the bound $|c_\mathrm{T}(z\simeq 0)/c-1|< 10^{-15}$~\citep{Baker:2017hug,Sakstein:2017xjx, Ezquiaga:2017ekz,Creminelli:2017sry}. 
This ruled out gravity models such as the quartic and quintic Galileons~\citep{Nicolis:2008in}, and placed restrictive constraints on others such as Horndeski theory~\citep{Horndeski:1974wa,DeFelice:2011hq}, TeVeS~\citep{Skordis:2009bf}, and Generalized Proca theory~\citep{Heisenberg:2017mzp}, if these models are invoked to explain cosmological observations. 
A sample of $\sim 10^2$ such multimessenger detections can provide strong constraints on any difference between the effective luminosity distance of GW sources and their electromagnetic counterparts~\citep{Belgacem:2019pkk}. 
This discrepancy in luminosity distances is another generic smoking gun of modified gravity that  could be used to test entire families of theories. 
The early warning and localization provided by a decihertz detector would be ideal for performing these multimessenger tests.

\subsection{Testing the Standard Model of particle physics through dark matter candidates}

One of the big mysteries in particle cosmology is that of dark matter: while overwhelming observational evidence ranging from flat rotation curves of galaxies to gravitational lensing indicates its existence, we are still in the dark when it comes to the constituents and properties of this elusive type of matter. 
BHs may come to our rescue, and serve as novel probes for axion-like particles that have become popular dark matter candidates~\citep{Hui:2016ltb,Arvanitaki:2009fg}.
This is possible because of the superradiant or BH bomb instability~\citep{Press:1972zz,Dolan:2007mj,Brito:2015oca}. 
Low frequency bosonic fields scattering off rotating BHs are superradiantly amplified and grow exponentially to form bosonic condensates if their Compton wavelength is comparable to the BH size. 

\begin{table}
\begin{center}
\caption{Relation between the black hole population, particle masses that we can search for (or constrain) and the relevant future gravitational-wave detector landscape.}
\label{tab:BHvsBSMParticles}
\begin{tabular}{ccl} 
    \br
    \multicolumn{1}{c}{Black hole mass $m_\mathrm{BH}/M_{\odot}$}    & Particle mass $m_\mathrm{B} c^2/\mathrm{eV}$ & \multicolumn{1}{c}{Detectors}     \\
    \mr
    $10^{9}$   & $10^{-21}$ & Pulsar timing array\\
    $10^{6}$   & $10^{-17}$ & \textit{LISA} \\
    $10^{3}$   & $10^{-14}$ & Decihertz \\
    $50$       & $10^{-12}$ & Ground-based    \\
    \br
\end{tabular}
\end{center}
\end{table}

%------------------------------------------
The latter implies that we can probe for a wide range of beyond-standard model particles from popular dark matter candidates to the quantum-chromodynamics axion using pure gravity, that more conventional
particle detectors cannot access, as illustrated in Table~\ref{tab:BHvsBSMParticles}. 
These yield important observable signatures such as gaps in the Regge (BH spin--mass) plane~\citep{Arvanitaki:2010sy, Pani:2012bp,Brito:2014wla,Ficarra:2018rfu}, 
and monochromatic GWs with frequencies (determined by the BHs' mass) across the spectrum~\citep{Witek:2012tr,Okawa:2014nda,Hannuksela:2018izj,Isi:2018pzk}. 
Similarly, compact binaries yield novel observational signatures due to resonances~\citep{Baumann:2018vus,Wong:2019yoc}. 
Hence, BHs and binaries thereof, and their GW emission can be used as innovative search engines for axion-like particles and beyond-standard model particles in general.

\subsection{Testing the physics of the early Universe and high-energy theories}

{GWs can carry unique information about the state of the Universe at epochs and energy scales far beyond the reach of current electromagnetic cosmological observables}. 
Information about the early Universe is encoded within the SGWB. 
The SGWB's has contributions from a variety of sources, including phenomena outside of the reach of electromagnetic probes. 
The SGWB's characteristic frequency today can be related to the Hubble factor at the generation time $H_*$~\citep{Caprini:2018mtu}, assuming that generation occurred during the radiation-dominated era:
\begin{equation}
\label{eq:fPT}
      f = 2.6 \times 10^{-8} \left(\frac{c k}{H_*}\right) \left(\frac{g_*(T_*)}{100}\right)^{1/6} \left(\frac{k_\mathrm{B} T_*}{\mathrm{GeV}}\right)~\mathrm{Hz},
\end{equation}
Here, $T_*$ is the Universe's temperature at the time the GW is sourced, $g_*(T_*)$ is the corresponding number of relativistic degrees of freedom, and $k$ is the wavenumber, such that the first term in parenthesis is the physical wavenumber at the time of GW production, normalised to the Hubble rate at that time (this factor depends on the details of the GW sourcing process, but for causality reasons must satisfy $c{k}/H_* \geq 1$).
The relationship between the temperature when the SGWB is generated and its characteristic frequency today illustrates how observing in different frequency bands probes GW emission from different epochs and energy scales in the early Universe. 
There are mainly two classes of SGWB source operating in the early Universe: those related to inflation and subsequent processes (such as reheating), and those related to primordial phase transitions. 

In all the below, we assume the absence of a SGWB of astrophysical origin which would mask the cosmological signal. 
The detection of an astrophysical SGWB, from stellar-mass binaries, is expected to be found and characterised by ground-based observatories~\citep{Abbott:2017xzg}. 
In the context of a DO, it would be necessary to accurately subtract out the astrophysical SGWB foregrounds~\citep[e.g.,][]{Cutler:2005qq,Pan:2019uyn,Pieroni:2020rob}; otherwise, the science of a cosmological SGWB would not be lost, but its effectiveness in constraining models for SGWB generation would be reduced. 
The improved knowledge of the rate of binary mergers provided by the large number of detections will make subtraction of the astrophysical SGWB easier. 

A SGWB is generically expected in the standard slow-roll inflationary scenario, extending in frequency with a slightly red-tilted spectrum from the horizon scale today to the one corresponding to the energy scale of inflation: $10^{-19}~\mathrm{Hz} <f<10^{11}~\mathrm{Hz}$~\citep{Caprini:2018mtu}.
Even though this signal intersects the frequency range of all GW detectors, measuring it is extremely challenging because of its low amplitude. 
At low frequencies $f<10^{-16}~\mathrm{Hz}$, the SGWB is the target of cosmic microwave background (CMB) experiments, through the measurement of the $B$-mode polarisation~\citep{Durrer:2008eom}. 
The present upper bound by the \textit{Planck} satellite on the tensor-to-scalar ratio is $r < 0.07$~\citep{Aghanim:2018eyx}, translating into a dimensionless energy density of $h^2 \Omega_\mathrm{GW} < 3 \times 10^{-16}$ (assuming no spectral tilt). 
This is expected to improve in the near future: on the ground, the Simons Array~\citep{Ade:2018sbj} could bound $r < 2 \times 10^{-3}$ by 2021--2025 and CMB Stage-IV~\citep{Abazajian:2016yjj} could reach $r < 10^{-3}$ by 2027--2031, and in space \textit{LiteBird}~\citep{Hazumi:2019lys} could reach  $r < 6 \times 10^{-4}$ by 2027--2032, and proposed satellites such as \textit{Pico}~\citep{Hanany:2019lle} or \textit{CORE}~\citep{Delabrouille:2017rct} could reach $r < 10^{-4}$, which is the lowest bound CMB experiments can technically reach. 
In the case of no positive detection by these CMB experiments, future direct GW detection would require sensitivity of $h^2 \Omega_\mathrm{GW} \sim 2 \times 10^{-19}$, corresponding to $r = 10^{-4}$, which is far below the sensitivity of any GW mission under study. 
However, there are scenarios, going beyond standard slow-roll inflation, in which the predicted SGWB spectral tilt becomes blue at high frequency, thereby opening up the possibility of a direct GW detection of the inflationary SGWB~\citep{Cook:2011hg,Bartolo:2016ami,Campeti:2020xwn}. 
This would constitute a major discovery, as it amounts to probing the inflationary potential near the end of inflation, which is observationally unconstrained. 
Consequently, it would provide crucial information about inflation and the high-energy physics model underlying it. 
Therefore, SGWBs from inflationary scenarios are an interesting, but speculative possibility; there are other sources of SGWBs which are more compelling for DOs.

The situation is different for sources in connection with primordial phase transitions. 
In particular, a first-order phase transition in the early Universe can generate a SGWB through the collisions of true-vacuum bubbles and the subsequent bulk motion of the plasma~\citep{Kosowsky:1992vn,Kamionkowski:1993fg,Gogoberidze:2007an,Caprini:2009yp,Hindmarsh:2013xza,Hindmarsh:2015qta}. 
In this case, the SGWB is expected to show a peak at a frequency scale set by the size of the bubbles when they collide, $R_* \sim v_\mathrm{w} /\beta$~\citep{Caprini:2015zlo}, where $v_\mathrm{w}$ is the bubble-wall speed and $1/\beta$ denotes the duration of the phase transition. 
Compared to \textit{LISA}, a DO would therefore be sensitive to first-order phase transitions occurring at higher temperature, cf.\ Eq.~\eqref{eq:fPT}, or with a shorter duration. 

The energy scale of the electroweak symmetry breaking (EWSB) corresponds to $T_*\sim 100~\mathrm{GeV}$; this phase transition offers a particularly interesting test-bed, since it certainly took place in the early Universe. 
In the context of the Standard Model of particle physics, it is a cross-over rather than a first-order transition~\citep{Kajantie:1995kf, Csikor:1998eu}; however, well-motivated scenarios beyond the Standard Model (BSM) predict a first-order EWSB, often together with baryogenesis processes or dark matter candidates~\citep{Delaunay:2007wb, Kozaczuk:2015owa, Kakizaki:2015wua, Chala:2016ykx, Dorsch:2016nrg, Bernon:2017jgv, Bruggisser:2018mrt, Chala:2018opy, YaserAyazi:2019caf}. 
Observing the SGWB signal from such a phase transition would therefore probe BSM physics at the $100~\mathrm{GeV}$ scale. 
Even if current and future colliders do not find BSM physics, there will be still room for a strong electroweak phase transition, as both the Future Circular Collider and the International Linear Collider would not be sensitive to the whole parameter space of the setups leading to the first-order electroweak phase transition~\citep{Fujii:2017vwa, Benedikt:2018csr,Caprini:2019egz}. 
In such a situation, a unique feature of a DO is its capability to probe phase transitions occurring at energy scales $T_*\sim 10$--$10^3~\mathrm{GeV}$ and lasting for a rather short time ($\beta/H_*\sim 10$--$10^6$).  
BSM descriptions of the EWSB typically predict weakly first-order (and consequently brief) phase transitions, with $10^2 \lesssim \beta/H_* \lesssim 10^4$~\citep{Caprini:2015zlo}. 
The corresponding SGWBs are outside the sensitivity range of \textit{LISA}, but could be detected by a higher frequency DO. 
Furthermore, weak phase transitions lead to sub-relativistic bubbles~\citep{Espinosa:2010hh, Megevand:2014yua, Huber:2013kj, Konstandin:2014zta, Dorsch:2018pat},
which increases the SGWB peak frequency. 
Therefore, a DO has the advantage of probing regions in the BSM EWSB parameter space which are more densely populated.

\begin{figure}[ht!]
\centering
 \includegraphics[scale=0.75]{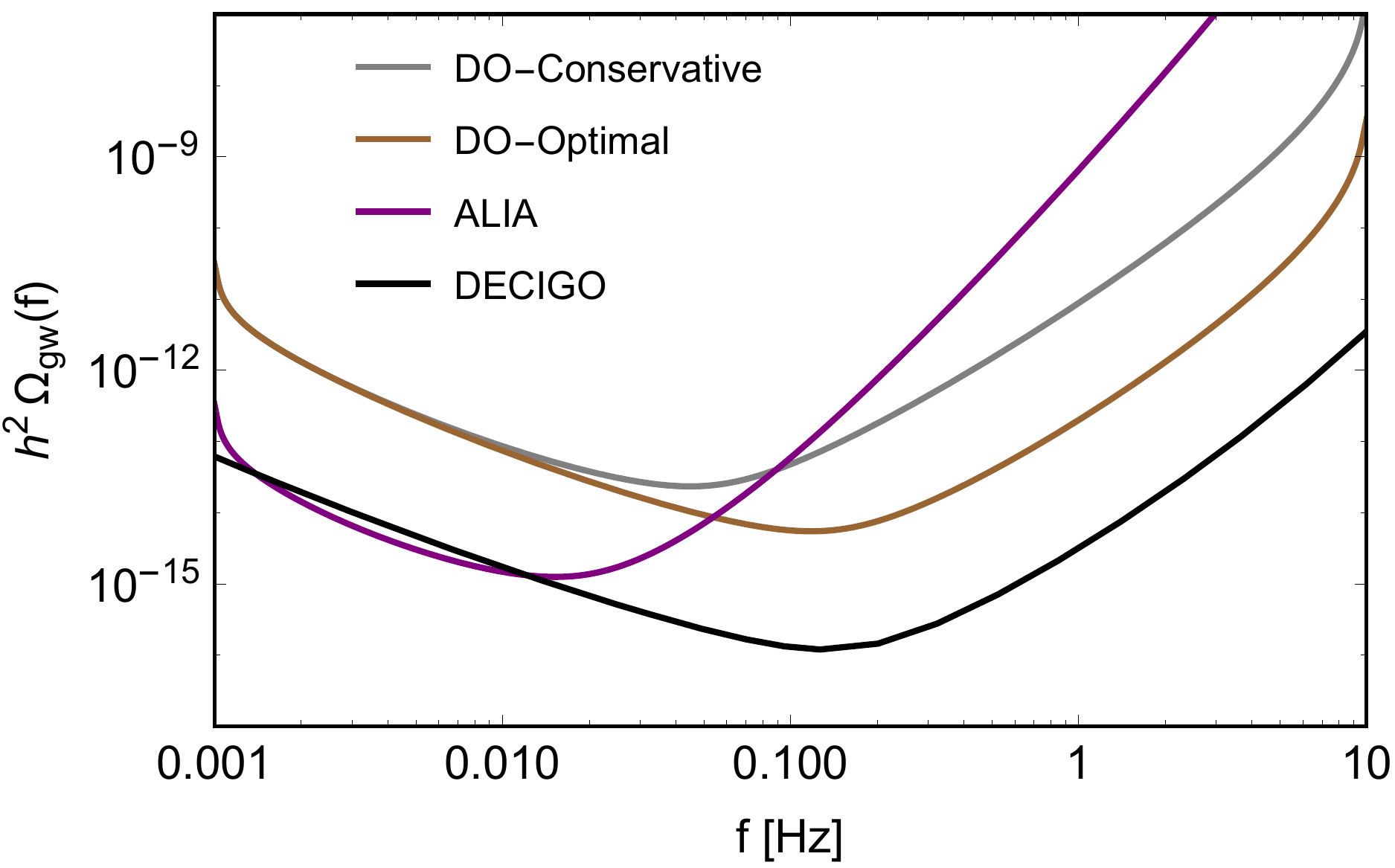} 
\caption{Ability to detect a stochastic gravitational-wave background, characterised by the power-law sensitivity $h^2 \Omega_\mathrm{GW}(f)$~\citep{Thrane:2013oya,Caprini:2019pxz,Mohamadnejad:2019vzg}. Sensitivities are calculated assuming a $4~\mathrm{yr}$ mission with a $70\%$ efficiency, and a required signal-to-noise ratio threshold of $10$.
}
\label{fig:PL}
\end{figure}

While DOs measure an promising frequency range, the amplitude of the SGWB decreases for weak and brief phase transitions. 
Therefore, to offer a realistic prospect of detecting the SGWB, the detector must have sufficient sensitivity. 
An instrument with $h^2 \Omega_\mathrm{GW} \sim 10^{-15}$ power-law sensitivity~\citep{Thrane:2013oya} around $0.1~\mathrm{Hz}$, such as the DO-Optimal or \textit{DECIGO} concepts (see Figure~\ref{fig:PL}), will be able to access interesting BSM scenarios predicting a first-order EWSB~\citep{Caprini:2015zlo}. 
For example, it could detect the SGWB from a phase transition featuring $k_\mathrm{B} T_* = 100~\mathrm{GeV}$, $\beta/H_*= 500$, $v_\mathrm{w} = 0.7 c$, $\alpha=0.07$, where $\alpha$ denotes the strength of the phase transition, given by the ratio between the vacuum energy density and the radiation energy density at the transition time. 
These are typical parameters in scenarios whereby the Standard Model is extended with an extra singlet~\citep{Espinosa:2011ax, Chen:2017qcz, Kozaczuk:2015owa} or embedded in supersymmetry~\citep{Huber:2015znp, Garcia-Pepin:2016hvs, Bian:2017wfv, Demidov:2017lzf}.  
Hence DOs have the sensitivity to discover SGWBs indicating BSM physics.

The SGWB signal from a source at $k_\mathrm{B} T_*\sim 1~\mathrm{TeV}$ and beyond, is also within the reach of a DO.
New physical phenomena around the $\mathrm{TeV}$ scale, possibly related to the presence of extra dimensions and leading to first-order phase transitions, have been widely studied in connection with the hierarchy problem or the presence of dark matter~\citep{Randall:2006py, Nardini:2007me, Konstandin:2010cd, Konstandin:2011dr, Bruggisser:2018mus, Megias:2018sxv}. 
The $100~\mathrm{TeV}$ scale emerges in new solutions to the hierarchy problem such as the relaxion~\citep{Arkani-Hamed:2015vfh, Graham:2015ifn}. 
Increasing $T_*$ far beyond the $\mathrm{TeV}$ scale corresponds to entering energy ranges for which there is no underlying theoretical physics model. 
This offers an amazing discovery potential. 

A primordial phase transition can also lead to the formation of topological defects. 
In BSM theories with extra symmetries (such as Grand Unified Theories), phase transitions are expected to occur, changing the symmetry of the vacuum. 
Stable defects can thereby form, depending on the details of the symmetry breaking scheme~\citep{Kibble:1976sj}.
Among these, are local (arising from the breaking of a local symmetry) cosmic strings which can be powerful SGWB sources~\citep{Vachaspati:1984gt}. 
Local cosmic strings arise naturally within well-motivated inflationary models (for instance, hybrid inflation~\citep{Jeannerot:2003qv}), but can also correspond to fundamental super-strings formed in brane inflation scenarios~\citep{Sarangi:2002yt}. 
The phase transitions creating cosmic strings can be first- or second-order, and they can be formed equally well during the thermal evolution of the Universe or during inflation.

Cosmic strings are characterized by the linear energy density $\mu$, which in the Nambu-Goto picture (describing infinitely thin string) is the tension.
This quantity is related to the energy scale of the phase transition, $\mathcal{E} \sim m_\mathrm{Pl}c^2 \sqrt{G\mu}$, where $m_\mathrm{Pl}$ is the Planck mass, and $G\mu$ is dimensionless. 
A (super-)string network permeating the Universe is formed both by infinite strings and string loops chopped-off when strings intersect or self-intersect. 
The loops oscillate relativistically and decay, releasing energy through GWs~\citep{Vachaspati:1984gt}.
Bursts of GWs are also produced by cusps, kinks and kink--kink collisions of cosmic strings~\citep{Damour:2001bk}. 
Even in the Nambu-Goto scenario, the evaluation of the SGWB from a string network is subject to several uncertainties. 
Adopting numerical-simulation results~\citep{Blanco-Pillado:2013qja}, the SGWB spectrum peaks at a frequency $f_\mathrm{peak}\propto (G\mu)^{-1}$~\citep{Blanco-Pillado:2017oxo}. 
Therefore, increasing the detector frequency opens up the possibility of constraining smaller values of $G\mu$, i.e.\ detecting the SGWB from cosmic strings produced at lower energy scales. 
However, the SGWB amplitude diminishes with the string tension (as $\sqrt{G\mu}$ in the radiation era plateau). 
Therefore, the detector sensitivity must also correspondingly improve to make this detection. 
The SGWB power-law sensitivity~\citep{Caprini:2019pxz} of \textit{LISA} goes down to $h^2\Omega_\mathrm{GW} \sim 10^{-13}$ and it will be able to probe string tensions down to $G\mu\sim 10^{-17}$~\citep{Auclair:2019wcv}, corresponding to energy scales $\mathcal{E} \sim 10^{10}~\mathrm{GeV}$. 
This already vastly improves present and future constraints from other GW detectors: the isotropic search for a SGWB during the first two LIGO--Virgo observing run set an upper bound at $G\mu \lesssim 10^{-6}$ in the most conservative scenario~\citep{LIGOScientific:2019vic,Abbott:2017mem}, while pulsar timing array observations constrain $G\mu \lesssim 10^{-11}$~\citep{Sanidas:2012ee,Blanco-Pillado:2017rnf}.
A GW observatory capable of reaching $h^2 \Omega_\mathrm{GW}\sim 10^{-15}$ at $0.1~\mathrm{Hz}$, such as the DO-Optimal or \textit{DECIGO} concepts (see Figure~\ref{fig:PL}), would be able to constrain the string tension down to $G\mu \sim 7 \times 10^{-20}$, corresponding to energy scales $\mathcal{E} \sim 10^{9}~\mathrm{GeV}$.

\section{Advancing technology}
\label{Sec:Detector}

\subsection{Detector technologies}

For the current generation of ground-based GW detectors---LIGO (USA and India), Virgo (Italy) and KAGRA (Japan)---the lowest accessible frequency is $\sim10~\mathrm{Hz}$~\citep{Aasi:2013wya}. 
This cutoff is primarily dictated by the seismic noise. 
To overcome this requires a new generation of ground-based GW detectors such as Cosmic Explorer (a $40~\mathrm{km}$ L-shaped interferometer)~\citep{Evans:2016mbw,Reitze:2019iox} or the Einstein Telescope (an underground $10~\mathrm{km}$ triangular detector) to access the frequencies down to $5~\mathrm{Hz}$ or $1~\mathrm{Hz}$ respectively~\citep{Punturo:2010zz,Sathyaprakash:2012jk}. 
This low-frequency sensitivity is vital for observing high-mass systems at cosmological distances: a $100 \Ms+ 100 \Ms$ binary at redshift $z = 10$ is not observable above $10~\mathrm{Hz}$~\citep{Kalogera:2019sui}.
The next-generation GW detectors are expected to commence operations in the 2030s, but will not push below $\sim1~\mathrm{Hz}$.

To break the $1~\mathrm{Hz}$ barrier requires a \emph{space-based} GW detector. 
The first generation of such observatories will be the European Space Agency's \textit{LISA} mission~\citep{Audley:2017drz}. 
\textit{LISA} consists of three satellites in a triangular constellation with an arm-length of $2.5\times10^9~\mathrm{m}$, trailing the Earth in a heliocentric orbit. 
It is primarily designed for peak sensitivity in the millihertz regime, but could in-principle detect GWs of frequency as high as $\sim0.1~\mathrm{Hz}$. 
The high-frequency sensitivity of \textit{LISA} is limited by the laser shot noise. 
There is hence a gap between space-based and ground-based detectors to be filled.

To probe GWs within the frequency spectrum of $\sim0.1$--$1~\mathrm{Hz}$ will require a space-based detector beyond \textit{LISA}~\citep{Baker:2019pnp}. 
Several proposals have been put forward by the global GW community for such a DO. 
One post-\textit{LISA} heliocentric mission concept is \textit{ALIA}~\citep{Bender:2013nsa,Baker:2019pnp}, which would have higher sensitivity than \textit{LISA} in $0.1$--$1~\mathrm{Hz}$ range. 
Other heliocentric DO concepts are \textit{Taiji}~\citep{Hu:2017mde}, which would have greatest sensitivity around $0.01~\mathrm{Hz}$, and \textit{TianGo}~\citep{Kuns:2019upi}, which would have greatest sensitivity in the $0.1$--$10~\mathrm{Hz}$ range.
Chinese scientists have pursued a geocentric mission concept \textit{TianQin}~\citep{Luo:2015ght} that would focus on continuous sources of GWs in the decihertz range, whilst having a range of other scientific targets~\citep{Wang:2019ryf,Shi:2019hqa,Liu:2020eko,Huang:2020rjf,Fan:2020zhy}.
A more ambitious concept design from Japan is \textit{DECIGO}~\citep{Sato:2017dkf,Kawamura:2020pcg}. 
This mission would have three individual $1000~\mathrm{km}$ interferometers, with arms consisting of Fabry--Perot cavities, in heliocentric orbit. 
The precursor to this mission, \textit{B-DECIGO}, would be a $100~\mathrm{km}$ triangular interferometer in the geocentric orbit. 
Both of these detectors would ensure a sensitivity many orders of magnitude better in the $0.1$--$1~\mathrm{Hz}$ range than \textit{LISA}. 
Scientists in the USA have proposed the \textit{Big Bang Observer} (\textit{BBO}), a concept which would essentially consist of four \textit{LISA} detectors in heliocentric orbits with combined peak sensitivity in the $0.1$--$1~\mathrm{Hz}$ range~\citep{Crowder:2005nr}.  
More modest designs are for geocentric constellations: the \textit{Geostationary Antenna for Disturbance-Free Laser Interferometry} (\textit{GADFLI})~\citep{McWilliams:2011ra} and \textit{geosynchronous Laser Interferometer Space Anetenna} (\textit{gLISA})~\citep{Tinto:2011nr,Tinto:2014eua} designs provide an order of magnitude improvement over \textit{LISA} in the decihertz range. 
More recently, there has been a proposal for a more cost-effective design, the \textit{SagnAc interferometer for Gravitational wavE} (\textit{SAGE})~\citep{Lacour:2018nws,Tino:2019tkb}, which consists of three identical CubeSats in geosynchronous orbit. 
Such CubeSat-based designs are promising in the decihertz regime, and their sensitivity can be improved by more powerful laser and better wavelength stabilization. 
These proposed designs are mostly variations on the classic \textit{LISA} design of a laser interferometer formed from a constellation of satellites.

In addition to technologies based on laser interferometry, new atom interferometer and atomic-clock-based approaches to DOs are in development. 
In these proposed atom-based approaches, the phase or frequency of a laser is differentially compared to atoms at both ends of a single baseline. 
The detected relative phase or frequency difference measures either GW induced effective changes in distance or effective Doppler shifts~\citep{Norcia:2017vwu}, while cancelling laser-frequency noise.
An appealing aspect of atom-based DOs is that they enable tuning of the detector transfer function by changing the laser sequences applied to the atoms locally, without requiring a corresponding change in spacecraft geometry~\citep{Graham:2016plp,Kolkowitz:2016wyg}. 
The range of technologies available mean that there are multiple possibilities for obtaining the necessary sensitivity in the decihertz range. 
The \textit{Mid-band Atomic Gravitational Wave Interferometric Sensor} (\textit{MAGIS})~\citep{Graham:2017pmn} and the \textit{Atomic Experiment for Dark Matter and Gravity Exploration in Space} (\textit{AEDGE})~\citep{Bertoldi:2019tck} are two DO concepts which use atom interferometry, in which ensembles of freely falling cold atoms measure the phase of a laser across a baseline consisting of two satellites in a geocentric orbit. 
Alternative schemes using optical lattice atomic clocks, in which the atoms in each of the satellites are strongly confined in an optical lattice referenced to a drag-free test mass, have also been proposed~\citep{Kolkowitz:2016wyg,Su:2017kng}. 
In Figure~\ref{Fig1}, the curve labeled Atomic Clock shows the projected strain sensitivity for one such optical-atomic-clock-based DO (based on \citep{Kolkowitz:2016wyg}, specifications given in Table~\ref{tab:design}). 
This level of differential clock stability has not yet been demonstrated terrestrially, but there is currently rapid improvement in both clock performance~\citep{Ludlow:2015,Campbell:2017,Oelker:2019} and atomic spin-squeezing~\citep{Hosten:2016,Braverman:2019,PedrozoPenafiel:2020}. 
Making optical clocks and matter-wave interferometers robust, portable, and space-hardy represents a significant additional technical hurdle to the realization of an atom-based DO. 
However, there has been successful creation of atomic Bose--Einstein condensates in sounding-rockets~\citep{Becker:2018} and on the \textit{International Space Station}~\citep{Aveline:2020kla}, and there are on-going efforts to realize portable optical clocks~\citep{Koller:2016hwj,Origlia:2018,Grotti:2018,Takamoto:2020tux} indicating that space-based versions may be achievable.
Concrete efforts to build terrestrial large-scale exploratory detectors (which will need to be further scaled up for GW science) are currently underway~\citep{Canuel:2017rrp,Coleman:2018ozp,Zhan:2019quq,Badurina:2019hst,Canuel:2019abg}.

\subsection{Illustrative designs}

\begin{table}
\begin{center}
\caption{Assumed technical specifications for Decihertz Observatory concepts. 
Acceleration and displacement noise are given relative to the \textit{LISA} values of $a_\mathit{LISA} = 3~\mathrm{fm\,s^{-2}\,Hz^{-1/2}}$ and $\Delta x_\mathit{LISA} = 10~\mathrm{pm\,Hz^{-1/2}}$ respectively. 
\textit{DECIGO} uses a cavity of finesse $10$~\citep{Sato:2017dkf}.}
\label{tab:design}
\begin{tabular}{l ccccc} 
    \br
    & \multicolumn{1}{c}{\textit{ALIA}} & \multicolumn{2}{c}{DO} & \multicolumn{1}{c}{\textit{DECIGO}} & \multicolumn{1}{c}{Atomic} \\
    &  & \multicolumn{1}{c}{Conservative} & \multicolumn{1}{c}{Optimal} &  & \multicolumn{1}{c}{Clock} \\
    \mr
    Arm length ($L/10^8~\mathrm{m}$)                       & $5$              & $1$              & $1$               & $10^{-2}$         & $15$              \\
    Acceleration noise ($a/a_\mathit{LISA}$)               & $10^{-1}$        & $10^{-1}$        & $10^{-1}$         & $10^{-4}$         & $10^{-4}$         \\
    Laser power ($P/\mathrm{w}$)                           & $30$             & $10$             & $30$              & $10$              & $1$               \\
    Telescope diameter ($D/\mathrm{m}$)                    & $1$              & $1$              & $2$               & $1$               & $1$               \\
    Laser wavelength ($\lambda/\mathrm{nm}$)               & $1064$           & $532$            & $532$             & $515$             & $689$             \\
    Displacement noise ($\Delta x/\Delta x_\mathit{LISA}$) & $5\times10^{-3}$ & $6\times10^{-4}$ & $8\times 10^{-5}$ & $2\times 10^{-6}$ & $4\times 10^{-2}$ \\
    Atoms per satellite ($N_\mathrm{a}$)                   & --               & --               & --                & --                & $10^9$            \\
    Atomic state squeezing ($S_\mathrm{a}/\mathrm{dB}$)    & --               & --               & --                & --                & $40$              \\
    Atom interrogation ($i_\mathrm{a}/\mathrm{s}$)         & --               & --               & --                & --                & $5$               \\ 
    Atom coherence time ($t_\mathrm{a}/\mathrm{s}$)        & --               & --               & --                & --                & $160$             \\
    \br
\end{tabular}
\end{center}
\end{table}

Throughout we have used two DO concept designs based on laser-interferometer technology, one moderately ambitious DO-Conservative and one more ambitious concept DO-Optimal. 
Together these designs bookend a plausible range for DOs.
By comparing the scientific potential of different designs, their relative merits can be assessed and weighed against the technological risks in achieving such engineering goals.
Given how well studied the \textit{LISA} design is, both DO-Conservative and DO-Optimal assume a triangular constellation in a heliocentric orbit for our illustrative concepts. 

\textit{LISA}-like mission concepts use some form of test masses that are ideally in perfect free fall and an interferometric readout to measure changes in the distance between these widely separated pairs of test masses.
The limiting noise sources for \textit{LISA}-like missions are typically separated into acceleration noise, or deviations from perfect free fall, and sensing noise in the interferometric readout system~\citep{Audley:2017drz,Cornish:2018dyw}. 
The former limits the performance at low frequencies, the later at high frequencies.
The sensitivities of the designs are shown in Figure~\ref{Fig1}.

For both DO-Conservative and DO-Optimal concepts we assume: $10^8~\mathrm{m}$ as a baseline, a factor $25$ shorter than for \textit{LISA} and a factor of $5$ shorter than \textit{ALIA}~\citep{Bender:2013nsa}, 
and a factor of $10$ improvement in acceleration noise beyond \textit{LISA Pathfinder} performance~\citep{Armano:2016bkm}. 
The main limiting noise sources for decihertz frequencies are force noise between the spacecraft and the test mass, and residual gas pressure. 

The force noise increases at high frequencies due to larger residual spacecraft motions; faster response times in the micronewton-thrusters would allow to reduce this residual motion. 
Better gravitational balancing, potentially in the form of actively controlled masses, would reduce the gravitational forces between spacecraft and test mass, and would also reduce the actuation noise. 
Residual gas pressure noise, from the Brownian motion of background gas, can be reduced by decreasing the pressure. 
\textit{LISA} requires that the gas pressure around the test mass is below $10^{-6}~\mathrm{Pa}$, and a similar standard should suffice to meet our DO concept designs.
Alone, these measures to reduce the force noise and gas pressure noise would allow for an order of magnitude improvement of the acceleration noise. 

The sensing noise is fundamentally limited by the laser shot noise. 
It depends on the received power, which is a function of the laser power, telescope diameter and laser wavelength. 
In terms of strain sensitivity, it is actually independent of the arm length; for longer arms, the received power drops quadratically due to diffraction which increases the phase noise and displacement sensitivity linearly. 
For DO-Conservative, the wavelength is changed to $532~\mathrm{nm}$ (from $1064~\mathrm{nm}$ used in \textit{LISA}), the laser power is increased by a factor of $5$ to $10~\mathrm{W}$, and the diameter of the telescope is expanded from $0.3~\mathrm{m}$ to $1~\mathrm{m}$.  
For DO-Optimal, these parameters are $532~\mathrm{nm}$, $30~\mathrm{W}$, and $2~\mathrm{m}$, respectively. 
Aside from shot noise, other subdominant sensing noise sources include thermal expansion and contraction of the telescopes, residual spacecraft motion which couples
to misalignments, laser frequency and intensity noise and timing noise. 
None of these are fundamental, and significant improvements will be possible. 
For example, one advantage of shorter arms in a heliocentric orbit are the reduced Doppler shifts between spacecraft. 
This would reduce the laser beat signals by a factor of $25$ compared to \textit{LISA}, and reduce the timing requirements by a similar factor. 
The advancements in detector technology are therefore challenging, but could be feasible for a DO mission in 2035--2050 given a concerted research-and-design effort.

\subsection{Conclusions}

Observing GWs in the decihertz range presents huge opportunities for advancing our understanding of both astrophysics and fundamental physics. 
The only prospect for decihertz observations is a space-based DO. 
Realising the rewards of these observations will require development of new detectors beyond \textit{LISA}. 
There is a wide range of potential DO mission designs, and we have illustrated the potential science returns of a selected few.
Meeting the technical requirements needed for a DO design will be challenging, but there are multiple promising approaches which could be developed to satisfy these goals.
A DO mission is potentially achievable within the coming decades, and such an endeavour would enable many new GW discoveries across astrophysics, cosmology and fundamental physics.

\ack{}
This paper is based upon a white paper submitted 4 August 2019 to ESA's Voyage 2050 planning cycle on behalf of the \textit{LISA} Consortium 2050 Task Force~\citep{Sedda:2019uro-v1}. 
Other space-based GW observatories proposed by the \textit{LISA} Consortium 2050 Task Force include a microhertz observatory $\mu$\textit{Ares}~\citep{Sesana:2019vho}; a more sensitive millihertz observatory, the \textit{Advanced Millihertz
Gravitational-wave Observatory} (\textit{AMIGO})~\citep{Baibhav:2019rsa}, and a high angular-resolution observatory consisting of multiple DOs~\citep{Baker:2019ync}.

The authors thanks Pete Bender for insightful comments.
MAS acknowledges financial support from the Alexander von Humboldt Foundation and the Deutsche Forschungsgemeinschaft (DFG, German Research Foundation) -- Project-ID 138713538 -- SFB 881 (``The Milky Way System''). 
CPLB is supported by the CIERA Board of Visitors Research Professorship. 
LS was supported by the National Natural Science Foundation of China (11975027, 11991053, 11721303) and the Young Elite Scientists Sponsorship Program by the China Association for Science and Technology (2018QNRC001).
TB is supported by The Royal Society (grant URF\textbackslash R1\textbackslash 180009). 
PAS acknowledges support from the Ram{\'o}n y Cajal Programme of the Ministry
of Economy, Industry and Competitiveness of Spain, as well as the COST Action
GWverse CA16104. This work was supported by the National Key R\&D Program of
China (2016YFA0400702) and the National Science Foundation of China (11721303).
EB is supported by NSF Grants No.\ PHY-1912550 and AST-1841358, NASA ATP Grants No.\ 17-ATP17-0225 and 19-ATP19-0051, NSF-XSEDE Grant No.\ PHY-090003, and by the Amaldi Research Center, funded by the MIUR program ``Dipartimento di Eccellenza''~(CUP: B81I18001170001). This work has received funding from the European Union’s Horizon 2020 research and innovation programme under the Marie Skłodowska-Curie grant agreement No.\ 690904. 
DD acknowledges financial support via the Emmy Noether Research Group funded by the German Research Foundation (DFG) under grant no. DO 1771/1-1 and the Eliteprogramme for Postdocs funded by the Baden-Wurttemberg Stiftung.
JME is supported by NASA through the NASA Hubble Fellowship grant HST-HF2-51435.001-A awarded by the Space Telescope Science Institute, which is operated by the Association of Universities for Research in Astronomy, Inc., for NASA, under contract NAS5-26555. 
MLK acknowledges support from the National Science Foundation under grant DGE-0948017 and the Chateaubriand Fellowship from the Office for Science \& Technology of the Embassy of France in the United States. 
IP acknowledges funding by Society in Science, The Branco Weiss Fellowship, administered by the ETH Zurich. 
AS is supported by the European Union's H2020 ERC Consolidator Grant ``Binary massive black hole astrophysics'' (grant agreement no.\ 818691 -- B Massive). 
NW is supported by a Royal Society--Science Foundation Ireland University Research Fellowship (grant UF160093).

\bibliographystyle{iopart-num}
\bibliography{biblio}

\end{document}